\documentclass[pra,showpacs,footinbib,amsmath,amssymb,twocolumn]{revtex4-1}

\usepackage{graphicx}
\usepackage{dcolumn}
\usepackage{bm}
\usepackage{amssymb}
\usepackage{amsmath}
\usepackage{times}
\usepackage{color}
\usepackage{setspace}
\usepackage{float}
\usepackage{subfigure}
\usepackage{hyperref}
\usepackage{braket}
\usepackage{mathtools}
\usepackage{refcount}
\usepackage{newfloat}
\usepackage{mathrsfs}
\DeclareFloatingEnvironment[name={S}]{suppfigure}

\allowdisplaybreaks
\begin{document}

\title{Evaporative cooling of reactive polar molecules confined in a
  two-dimensional geometry\\}

\author{Bihui Zhu$^{1}$, Goulven Qu\'em\'ener$^2$, Ana M. Rey$^1$} \email{arey@jilau1.colorado.edu}

\author{Murray J. Holland$^1$} \email{murray.holland@colorado.edu}

\affiliation{$^{1}$JILA and Department of Physics, University of
  Colorado, 440 UCB, Boulder, CO 80309, USA }%

\affiliation{$^2$ Laboratorie Aim\'e Cotton, CNRS, Universit\'e
  Paris-Sud, ENS Cachan, Campus d'Orsay, B\^at. 505, 91405 Orsay,
  FRANCE}

\date{\today}

\begin{abstract}

  Recent experimental developments in the loading of ultracold KRb
  molecules into quasi-two-dimensional traps, combined with the
  ability to tune the ratio between elastic and loss
  (inelastic/reactive) collisions through application of an external
  electric field, are opening the door to achieving efficient
  evaporative cooling of reactive polar molecules. In this paper, we
  use Monte Carlo simulations and semianalytic models to study
  theoretically the experimental parameter regimes in which
  evaporative cooling is feasible under current trapping
  conditions. We investigate the effect of the anisotropic character
  of dipole-dipole collisions and reduced dimensionality on
  evaporative cooling. We also present an analysis of the
  experimentally relevant anti-evaporation effects that are induced by
  chemical reactions that take place when more than one axial
  vibrational state is populated.

\end{abstract}

\pacs{34.50.Cx, 37.10.Mn, 05.20.Dd}
\maketitle

\section{\label{sec:1}Introduction}
Polar molecules that exhibit strong dipole-dipole interactions provide
a flexible platform for realizing a broad range of interesting
phenomena relevant to condensed matter physics, quantum information
sciences and precision
measurements~\cite{lahayephysics,precision1,precision2,moleculeQM,
  mQM, moleculetoolbox, carrcold}. The parameter regime of interest is
generally ultralow temperature and high phase-space density, where
novel quantum features can emerge. In practice, it turns out to be
difficult to cool molecules into the desired regime using standard
methods due to their complicated internal level structure. In the past
few years, significant experimental progress has been made towards
this goal through the demonstration of a method for preparing a gas of
fermionic KRb molecules in the lowest electronic, vibrational, and
rotational quantum state, with a temperature $T$ at the verge of
quantum degeneracy ({\em i.e.}~$T/T_{\rm{F}}\sim 1$ where $T_{\rm F}$
is the Fermi temperature)~\cite{ni2008, coldpolar2,ninature}. From
this starting point, one would like to further increase the
phase-space density by implementing evaporative cooling---demonstrated
to be one of the most useful cooling methods for quantum
gases~\cite{evap1,evap2,evap3}.

A fundamental limitation to the effectiveness of evaporative cooling
for polar molecules is their fast losses. For KRb molecules, the
losses arise mainly from exothermic chemical reactions, {\em i.e.}
KRb~+~KRb$\rightarrow$ K$_2$~+~Rb$_2$. In such reactions, molecules
prepared in different internal states can undergo barrierless
collisions, with a lifetime of only $\sim$10~ms~\cite{ninature,
  goulvenloss}. In contrast, identical fermionic KRb molecules at
ultra-low temperature are protected by the $p$-wave barrier, which
potentially results in a much slower reaction
rate~\cite{ninature}. However, due to the anisotropic nature of the
dipole-dipole interaction, molecules of the correct orientation can be
attracted towards each other by experiencing ``head-to-tail''
collisions. As a result, the $p$-wave barrier can be lowered by the
application of a strong external field, increasing the loss mechanism
of the molecules.

Significantly, these obstacles can be overcome by confining the
molecules into quasi-two-dimensional traps, which can be generated by
a one-dimensional optical lattice. In this case, the adverse
collisions can be greatly suppressed and the reaction barrier
effectively raised~\cite{stereoloss, goulvenband,
  goulven2d}. Furthermore, both the elastic and reactive collision
rates can be tuned by controlling the applied external electric field
and the trapping potential. Thus, while most previous experiments
implementing evaporative cooling were performed in three-dimensional
geometries~\cite{ketterlereview}, here we are explicitly interested in
focussing on evaporative cooling in two-dimensions, with the
anisotropic collisions that arise from the dipolar interaction. A
detailed understanding of this situation would be beneficial for
future experimental realizations.

Given this motivation, we theoretically investigate evaporative
cooling of molecules. We use both Monte Carlo (MC) simulations and
models developed on the basis of kinetic theory to study the
efficiency of evaporative cooling with parameters applicable to current
state-of-the-art KRb experiments. The paper is structured as
follows. In Sec.~\ref{sec:aniso} we explore the effect of anisotropic
collisions on evaporative cooling. In Sec.~\ref{sec:2dvs3d} we
consider the effect of reduced trap dimension, {\em
  i.e.}~two-dimensional rather than three-dimensional traps. In
Sec.~\ref{sec:KRb} we apply the MC method to determine the optimum
evaporative cooling trajectory for KRb molecules. In
Sec.~\ref{sec:antievap} we discuss the potential anti-evaporation
mechanism arising from the energy dependence of the reactive loss.

\section{\label{sec:aniso} Evaporative cooling with anisotropic
  collisions}
\subsection{\label{subsec:sigma} Anisotropic elastic collisions
  between polar molecules}

Evaporative cooling relies on removing particles with above-average
energy and redistributing the residual energy among the remaining
particles by elastic collisions so that the temperature falls. For
polar molecules, the characteristic parameters that encapsulate the
elastic scattering process can be dramatically modified by the
application of external fields. In the presence of an external
electric field, the dipole-dipole interaction between polar molecules
can mix different partial waves and give rise to highly-anisotropic
scattering. Moreover, for identical fermions, the lowest total angular
momentum partial wave that has the correct symmetry is $p$-wave, so
the elastic collisions are anisotropic at ultralow temperature even in
the absence of applied fields~\cite{goulvendipole}.

Considering current experimental conditions~\cite{stereoloss}, we have
computed the differential and total scattering cross-section for KRb
molecules in two-dimensions at an induced dipole moment $d=0.2$~debye
as a function of the collision energy $E_{\rm{c}}$. The theoretical formalism is explained in Appendix~\ref{append:cross}. FIG.~\ref{fig:kappa}(a) shows the dependence of differential
cross-section on the scattering angle, $\phi$, which is the precession
angle of the relative momentum during the collision. Scattering is
mainly forward ($0$ or $2\pi$) and backward ($\pi$) after a collision. The peak is more
pronounced as the collision energy increases due to the larger
contribution of higher partial waves.  This differential cross-section
is well parametrized by the empirical function
\begin{equation}
\left(\frac{d\lambda(\phi)}{d\phi}\right)_{E_{\rm{c}}}
=\lambda(E_{\rm{c}}) \left(a(\text{cos}\phi)^{2\alpha}
+a'(\text{cos}\phi)^{2\alpha'}\right),\label{eq:diffsigma}
\end{equation}
with $a$, $a'$, $\alpha$ and $\alpha'$ constants that are real and
positive, and with the total cross-section~\cite{fnote} given by
\begin{equation}
  \lambda(E_{\rm{c}})=\int_0^{2\pi}d\phi\left(\frac{d\lambda(\phi)}{d\phi}\right)_{E_{\rm{c}}}.
\end{equation}
The dependence of $\lambda$ on collision energy $E_{\rm{c}}$ is shown in
FIG.~\ref{fig:kappa}(b). Our best parametrization for various scattering energies is
given in Table~\ref{table1}.  Note that at $E_{\rm{c}}\sim1$~nK, the
angular dependence of the differential cross-section is well described by that of the lowest odd partial wave (see Appendix.~\ref{append:cross}), since $a\gg a'$ and $\alpha\approx1$.

\begin{table}
\centering
\setlength{\tabcolsep}{6pt}
\begin{tabular}{|r|r|r|r|r|c|}\hline
  $E_{\rm{c}} (\rm{nK})$&$a$&$a'$&$\alpha$&$\alpha'$&$\lambda (10^{-6}\rm{cm})$\\\hline
  1 &0.31 &0.0005&1.00&2.19&0.38\\\hline
  10 &0.27 &0.06&1.00&2.09&3.67\\\hline
  100 &0.24 &0.21&1.19&7.03&5.99\\\hline
  1000&0.34 &0.40&2.47&26.40&3.42\\\hline
\end{tabular}
\caption{\label{table1}Derived parameters from the fit of 
  Eq.~(\ref{eq:diffsigma}) for the scattering energies shown 
  in FIG.~\ref{fig:kappa}.}
\end{table}

\begin{figure*}
  \begin{center}
  \includegraphics[width=0.99\textwidth]{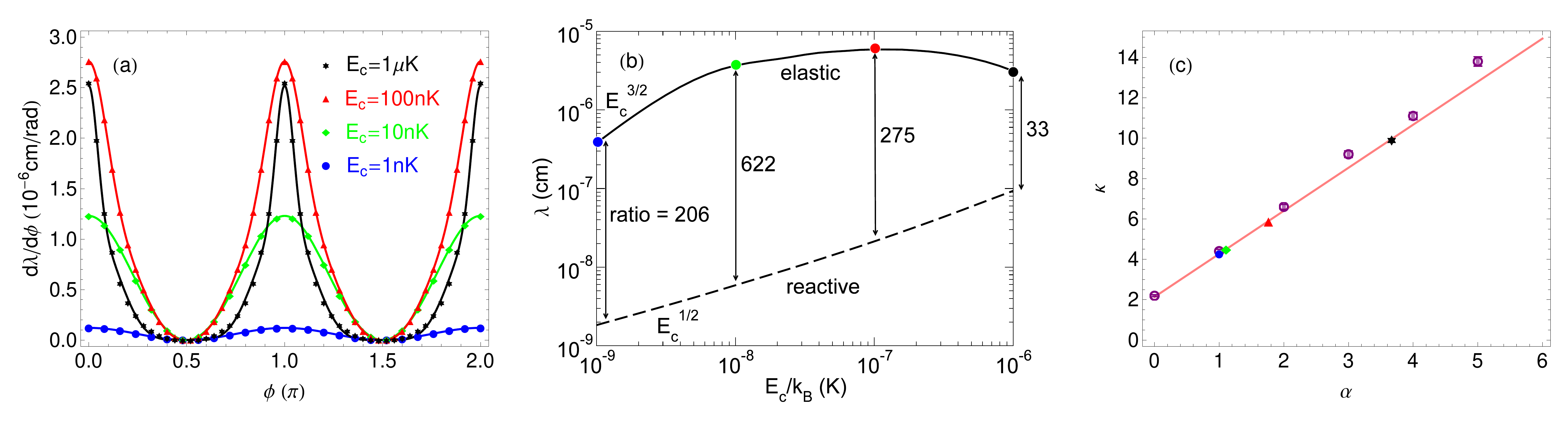}
 \end{center}
\caption{(Color online) \label{fig:kappa} (a)~Two-dimensional
  differential scattering cross-section for elastic collisions as a
  function of the scattering angle. Four collision energies are shown;
  $E_{\rm{c}}=1$~nK (blue disk), 10~nK (green diamond), $100$~nK (red
  triangle), and 1~$\mu$K (black star). The induced dipole moment is
  $d=0.2$~debye and the harmonic frequency of the one-dimensional
  confinement potential is $\nu=23$~kHz, The solid lines show the
  empirical formula Eq.~(\ref{eq:diffsigma}). (b)~Total scattering cross-section for elastic collisions (solid line) and reactive processes
    (dashed line) as a function of the collision energy for an induced
    dipole moment of $d=0.2$~debye and a one dimensional confinement
    of $\nu=23$~kHz. The ratio of elastic to reactive is indicated at
    four different collision energies; $E_{{\rm c}}=$1~nK (blue point),
    10~nK (green point), 100~nK (red point), and 1~$\mu$K (black
    point). (c)~Number of collisions
  $\kappa$ required for thermalization for different differential
  scattering cross-sections. The magenta solid line shows
  Eq.~(\ref{eq:kappa}). The purple circles are results from the MC
  simulations, and are shown with their small statistical error
  bars. The four symbols (blue disk, green diamond, red triangle,
  black star) correspond to the same energies as (a).
}\label{fig:zenoex}
\end{figure*}
\subsection{\label{subsec:therm}Thermalization rate}

The thermalization rate characterizes the timescale needed for a
system to redistribute energy after an evaporative cut. To
quantitatively investigate the thermalization under anisotropic
collisions, we adopt the typical experimental procedure of
cross-dimensional thermalization~\cite{thermcross, pwaveMC}. In order
to isolate the effect of anisotropic elastic collisions, we will
neglect losses completely, and also neglect the energy dependence in the total
elastic cross-section $\lambda$ for these thermalization calculations.

Consider a gas of $N$ molecules in a two-dimensional harmonic trap
with a slight initial imbalance of temperatures along each axis. This
initial condition can be prepared in experiment by parametric heating
of an equilibrium gas~\cite{thermcross}. We define an effective
temperature $T_i$ in the $i$th direction in terms of the total energy $E_i$
by $k_BT_i=E_i/N=
\overline{p_i^2}/(2m)+m\omega_i^2\overline{x_i^2}/2$, where $\omega_i$
is the trapping frequency, $m$ is the molecule mass, and
$\overline{p_i^2}/(2m)$ and $m\omega_i^2\overline{x_i^2}$ denote the
average kinetic energy and potential energy. We assume that the system
is well described by the Boltzmann distribution
\begin{equation}
  f_0(\mathbf{x},\mathbf{p})=n_0\exp\Bigl(-\sum_{i}{(\frac{p_i^2}{m}
    +m\omega_i^2x_i^2)/(2k_BT_i)}\Bigr),
  \label{eq:bolt}
\end{equation}
where $n_0$ guarantees normalization, {\em i.e.}
\begin{equation}
  \frac{1}{(2\pi\hbar)^2}\int
  d^2xd^2p\,f_0(\mathbf{x},\mathbf{p})=N
  \,.
\end{equation}
Without loss of generality, we assume $T_y=(1-\xi)T_x$ with
$\xi>0$. Elastic collisions lead to an exchange of energy between the
$x$ and $y$ directions and reduce the relative temperature
difference. The rate of such change is~\cite{enskog}
\begin{eqnarray}
  \frac{d E_x}{dt}&=& N k_B\frac{d T_x}{dt} 
  =\frac{1}{2m(2\pi \hbar)^4}\int d^2x
  d^2p_1d^2p_2d\phi' \nonumber\\
  &&{}\times f_0(\mathbf{x},\mathbf{p}_1)f_0(\mathbf{x},\mathbf{p}_2)
  |\mathbf{p}_1-\mathbf{p}_2|\frac{d\lambda(\phi')}{d\phi'}
  \Delta E_x .\label{eq:dEx}
\end{eqnarray}
This involves the energy change per collision given by
\begin{equation}
\Delta E_x=\frac{1}{4m}|\mathbf{p}_1-\mathbf{p}_2|^2
(\text{cos}^2\phi'-\text{cos}^2\phi),
\end{equation}
where $\phi$ and $\phi'$ are the angles between the relative momentum
and the total momentum before and after the collision respectively.
The period of time it takes to thermalize, $\tau$, can be defined as
the $1/e$ decay time of the temperature difference, so that
\begin{eqnarray}
\frac{d T_x}{dt}=&&-\frac{1}{\tau}(T_x-T_0),\label{eq:dTx}
\end{eqnarray}
where $T_0=(T_x+T_y)/2$ is the temperature at equilibrium. From
Eq.~(\ref{eq:dEx}) and Eq.~(\ref{eq:dTx}), we obtain the average
number of elastic collisions required for the system to thermalize,
$\kappa$, in the following way. We begin by defining the collision
rate $\gamma$
\begin{equation}
  \gamma=\frac{\lambda}{m(2\pi\hbar)^4}
  \int d^2xd^2p_1d^2p_2\,f_0(\mathbf{x},\mathbf{p}_1)
  f_0(\mathbf{x},\mathbf{p}_2)|\mathbf{p}_1-\mathbf{p}_2|,
\label{eq:gamma1}
\end{equation}
so that $\kappa\equiv\lim_{\xi\rightarrow0}(\tau\gamma)$, independent
of the initial temperature and trapping frequencies~\cite{pwaveMC,
  jacobscatter}. Combining Eq.~(\ref{eq:bolt})--Eq.~(\ref{eq:gamma1})
with the form of the differential cross-sections given in
Eq.~(\ref{eq:diffsigma}) leads to the following result:
\begin{eqnarray}
  \kappa=&&\frac{8}{15\sqrt{\pi}}\bigg\{\frac{a}{2\alpha+2}
  \frac{\Gamma\left(\alpha+\frac{1}{2}\right)}{\Gamma(\alpha+1)}+\frac{a'}{2\alpha'+2}
  \frac{\Gamma\left(\alpha'+\frac{1}{2}\right)}{\Gamma(\alpha'+1)}
  \bigg\}^{-1}.\label{eq:kappafull}\nonumber\\
\end{eqnarray}
When $a'=0$ and $\alpha'=0$, this reduces to the particularly simple
expression
\begin{equation}
\kappa=\frac{16}{15}(2\alpha+2), \label{eq:kappa}
\end{equation}
which corresponds to the differential cross-section
\begin{equation}
\frac{d\lambda(\phi)}{d\phi}=a~\text{cos}^{2\alpha}\phi.\label{intform}
\end{equation}
In order to interpret these results, for each scattering energy~$E_{\rm{c}}$,
we calculate $\kappa$ using Eq.~(\ref{eq:kappafull}), and plot the
result on the $(\alpha,\kappa)$ line, as prescribed by
Eq.~(\ref{eq:kappa}), with markers (star, square, triangle, diamond,
and disk), as shown in FIG.~\ref{fig:kappa}(c). The use of the linear
relationship between $\alpha$ and~$\kappa$ is supported by results
obtained from MC simulations, as described in Appendix~\ref{sec:MC},
utilizing the differential cross-section form of Eq.~(\ref{intform})
directly (purple circles). The exception is at large $\alpha$, where
the assumption adopted in deriving Eq.~(\ref{eq:kappafull})---namely
that of a Boltzmann distribution parametrized by an effective
temperature---becomes poor when the thermalization is too slow.

The calculations demonstrate a strikingly strong dependence on the
anisotropy of the collisions for rethermalization. For $\alpha=0$ the
collisions are isotropic and $\kappa\approx2.1$. In comparison, $\kappa\approx 4.3$ at $\alpha=1$, which describes the lowest partial wave scattering between identical fermions. For~$E_{\rm c}=1\,{\mu\rm K}$ we find
$\kappa\approx9.7$, which is many times that for isotropic collisions,
implying a significant increase in the number of collisions required
for rethermalization. One may have anticipated this since the
differential cross-section is more and more sharply peaked as $\alpha$ increases, and the energy is less and less efficiently distributed. Also
note that the outcome for the differential cross-section of KRb with
$E_{\rm{c}}=$1~nK is very close to that from Eq.~(\ref{eq:kappa}) with
$\alpha=1$, indicating that the lowest odd partial wave dominates for elastic collisions between fermions in indistinguishable internal and external state at ultra-low temperature.

\subsection{\label{subsec:anisoevap}Evaporation with anisotropic
  collisions}

\begin{figure}\centering
\includegraphics[width=0.42\textwidth]{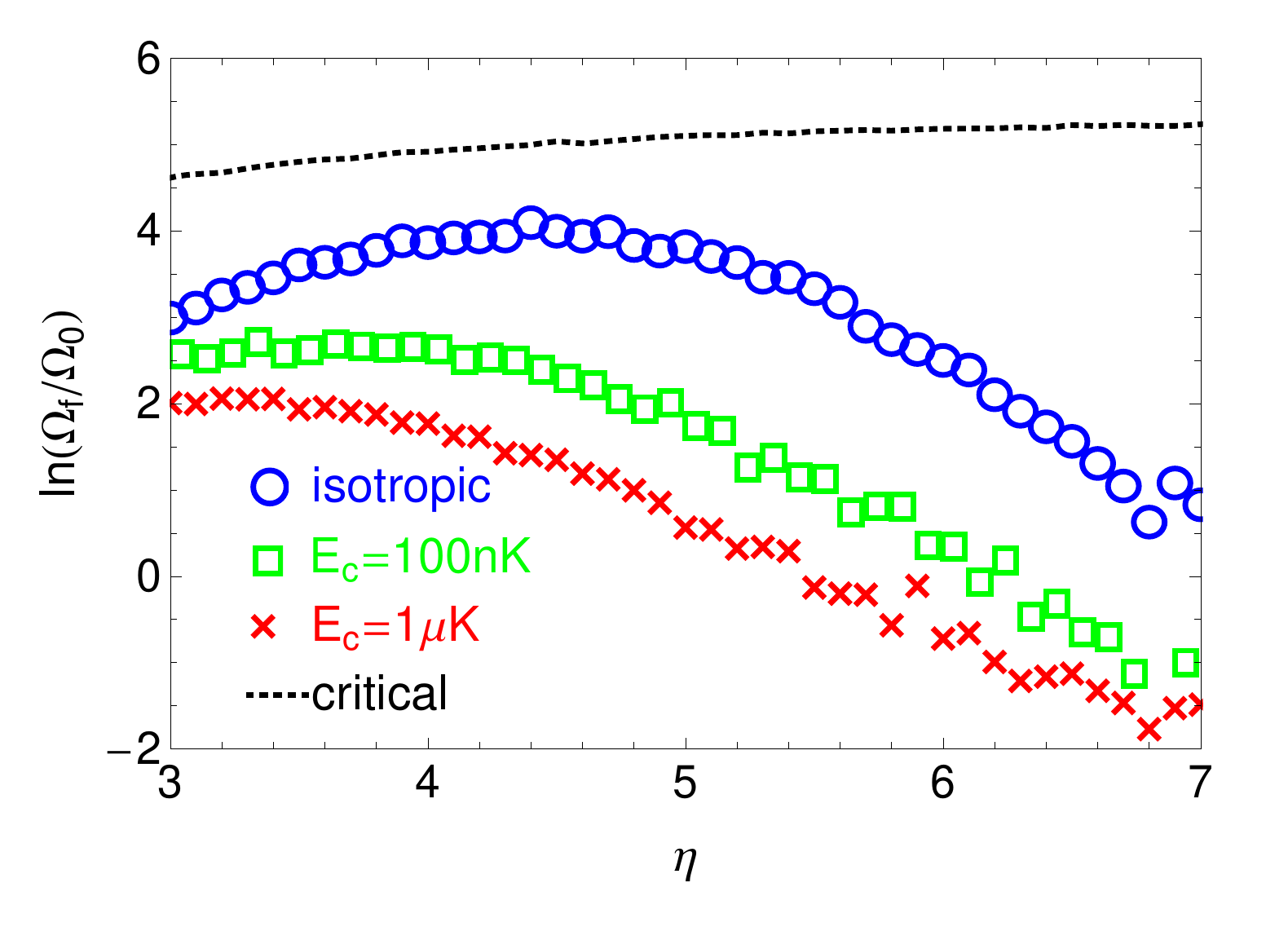}
\caption{\label{fig:pdia}(Color online) Increase in phase space
  density vs truncation parameter $\eta$ for different
  $\frac{d\lambda(\phi)}{d\phi}$. Here $\Omega_0$ is the initial
  phase-space density, and the final phase space density,
  $\Omega_{\rm{f}}$, was calculated when the ratio of molecule numbers
  was $N_{\rm{f}}/N_0=0.1$.  Blue circles:
  $\frac{d\lambda(\phi)}{d\phi}=$const; red crosses:
  $(\frac{d\lambda(\phi)}{d\phi})_{E_{\rm{c}}=1\mu\mathrm{K}}$; green squares:
  $(\frac{d\lambda(\phi)}{d\phi})_{E_{\rm{c}}=100\,{\rm nK}}$; black dotted
  line: critical phase space density $\Omega_c$ for quantum
  degeneracy. }.
\end{figure}

\begin{figure}\centering
  \includegraphics[width=0.495\textwidth]{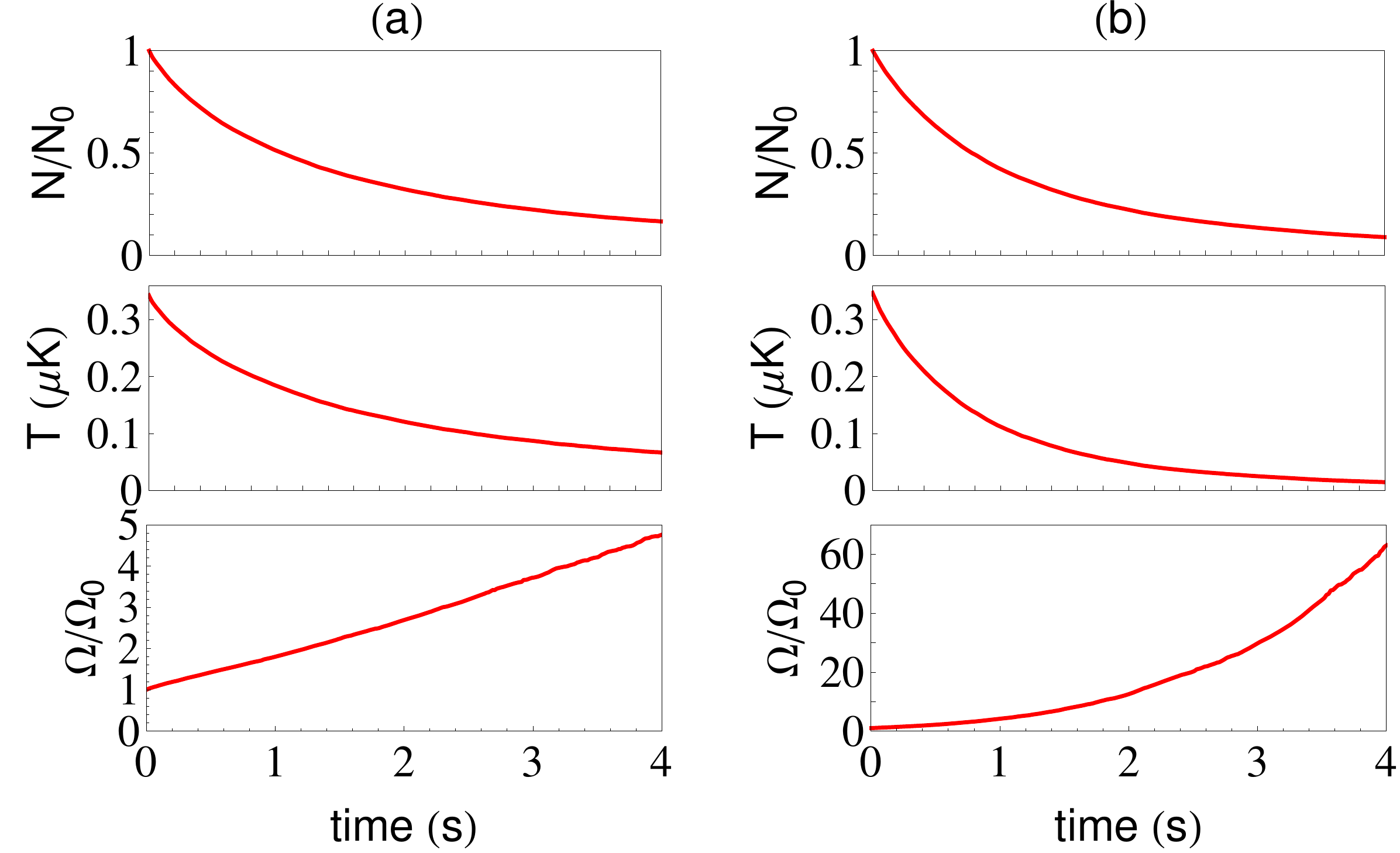} 
  \caption{\label{fig:A800}(Color online) Trajectories of evaporation for (a):
    $(\frac{d\lambda(\phi)}{d\phi})_{E_{\rm{c}}=1\mu \mathrm {K}}$ at
    $\eta= 3.3$ and (b): $\frac{d\lambda(\phi)}{d\phi}=$const at
    $\eta=4.3$, under the same initial conditions as in
    FIG. \ref{fig:pdia}. In comparison with (a), it is apparent that
    in (b) the temperature $T$ drops faster and the phase-space
    density $\Omega$ increases much more rapidly.}
\end{figure}

In evaporative cooling experiments, loss collisions compete with
elastic collisions, and thus a slow thermalization rate gives rise to
a reduction in cooling efficiency.  In this section, we use MC
simulations to investigate the efficiency of evaporative cooling in
the presence of anisotropic elastic collisions, and also including two-body reactive collisions, as occur for KRb
molecules. We quantify the efficiency in two ways: by the achieved
increase in phase-space density $\Omega_{\rm{f}}/\Omega_0$ at the
expense of losing a certain portion of the molecules in the trap (see
FIG.~\ref{fig:pdia}), and by the time required for the temperature to
go down and $\Omega$ to increase by a given amount (see
FIG.~\ref{fig:A800}).

In these simulations, we assume an instantaneous removal of all
particles with energy greater than a cut-off energy. The cut-off
energy, $\epsilon_t$, evolves during the evaporation trajectory with
the constraint that a truncation parameter,
$\eta=\epsilon_t/\overline{E}$, is kept constant, where $\overline{E}$
is the average energy of the nonequilibrium distribution. The
comparison between isotropic and anisotropic collisions is made by
calculating the evaporation trajectory for a variety of differential
cross-sections, starting with the same initial molecule number. To
simplify the comparison, for both elastic and reactive collisions, we
keep the total cross-sections constant and energy independent. More specifically, we consider elastic differential cross-sections in the form of Eq.~(\ref{eq:diffsigma}), and choose the same $\lambda(E_{\rm{c}})=\lambda_{el}$  for all differential cross-sections used in our simulations, leaving $a,a',\alpha$ and $\alpha'$ defined as in Table~\ref{table1} ($a=1, a'=\alpha=\alpha'=0$ for isotropic collisions). And we fix the reactive cross-section via $\lambda_{re}=\zeta\lambda_{el}$, with $\zeta$ a constant ratio.  

The molecules are initially simulated from a truncated Boltzmann
distribution utilizing the cut-off energy~\cite{walraven}
$\epsilon_t$:
\begin{eqnarray}\label{eq:tbf}
  f(\mathbf{x},\mathbf{p})=&&n'_0\exp\bigl(-(\frac{p^2}{m}
  +m\omega^2x^2)/(2k_BT)\bigr)\nonumber\\
  &&\times\Theta\bigl(\epsilon_t-(\frac{p^2}{2m}
  +\frac{1}{2}m\omega^2x^2)\bigr),
\end{eqnarray}
where $\Theta$ is the Heaviside function, $n'_0$ is the normalization,
and $T$ is the initial temperature, held fixed for different
simulations.  Although the energy distribution of the molecules is
intrinsically in nonequilibrium here, we may assign it a phase-space
density, $\Omega$, as that of the corresponding equilibrium
distribution that has the same molecule number and average molecule
energy.

In FIG.~\ref{fig:pdia}, $\Omega$ is plotted for the case of
evaporation trajectories in which 90\% of the initial molecules are
lost. For anisotropic collisions, we calculate evaporation
trajectories for the differential cross-sections for $E_{\rm
  c}=100\,{\rm nK}$ and $E_{\rm c}=1\mu\mathrm {K}$ given in the
subsection~\ref{subsec:therm}. The results presented in
FIG.~\ref{fig:pdia} show that the maximum efficiency achievable by
varying $\eta$ is much lower for anisotropic collisions than that
achievable with isotropic collisions. In addition, the maximum
efficiency is reached at smaller $\eta$. This suggests, for example,
that if $\epsilon_t$ is set by the trap depth in an experiment, a
shallower trap is desirable for evaporating a gas with anisotropic
collisions, compared with evaporating a gas with isotropic collisions,
at the same temperature and collision rate.

For these calculations, the reactive to elastic collision rate ratio
was held fixed at $\zeta=1/200$. This was chosen based on typical
experimental conditions for KRb molecules (see
Sec.~\ref{sec:KRb}). The value of $\zeta$ does not qualitatively
change the above comparisons; it only changes the shape of each
curve~\cite{ketterlereview}, and as one would expect, the maximum
efficiency and the corresponding $\eta$ decreases as $\zeta$
increases.

In FIG.~\ref{fig:A800}, we plot the time-dependent trajectories of
evaporative cooling for isotropic collisions with $\eta= 4.3$, and for
anisotropic collisions for $E_{\rm c}=1~\mu\mathrm {K}$ with
$\eta= 3.3$. These values of $\eta$ were chosen such that
$\Omega_{\mathbf{f}}/\Omega_0$ is maximum in FIG.~\ref{fig:pdia}, and
thus most efficient. It is evident from this plot that to reach the
same increase in $\Omega$ and decrease in $T$, it takes much shorter
time with isotropic collisions (FIG.~\ref{fig:A800}(b)). This
indicates a lower efficiency with anisotropic collisions, consistent
with the results of FIG.~\ref{fig:pdia}.

\section{\label{sec:2dvs3d}Evaporative cooling in 2D traps}

Physics in two-dimensional geometries can often be quite different
from that in three-dimensions. Unlike many previous evaporative
cooling experiments, KRb molecules need, in general, to be confined in
traps with reduced dimensionality to be stable. Such a configuration
was also used to evaporate Cs atoms~\cite{chin2D}. Here, we study the
effect of reduced dimensionality on evaporative cooling by comparing
the efficiency in two-dimensional and three-dimensional traps.

The previous section showed that the existence of highly-anisotropic
collisions slows down thermalization and decreases the efficiency of
evaporative cooling. A calculation analogous to
Sec. \ref{subsec:therm} but for a three-dimensional harmonic trap
gives
\begin{equation}
\kappa=\frac{5}{6}(2\alpha+3),
\label{eq:kappa3d}
\end{equation}
which implies $\kappa=2.5$ when $\alpha=0$ (isotropic
collisions). This result may suggest a potentially faster
thermalization rate and more efficient evaporation in a
two-dimensional harmonic trap when compared with a three-dimensional
harmonic trap with everything else held fixed. However, such a
conclusion based solely on Eq.~(\ref{eq:kappa3d}) would be premature,
since the density of states depends on the
dimensionality. Specifically, it was found for the comparison between
quadratic potentials and linear potentials in three dimensions that
changing the density of states can also change the evaporative cooling
efficiency~\cite{ketterlereview}.

\begin{figure}\centering
  \includegraphics[width=0.42\textwidth]{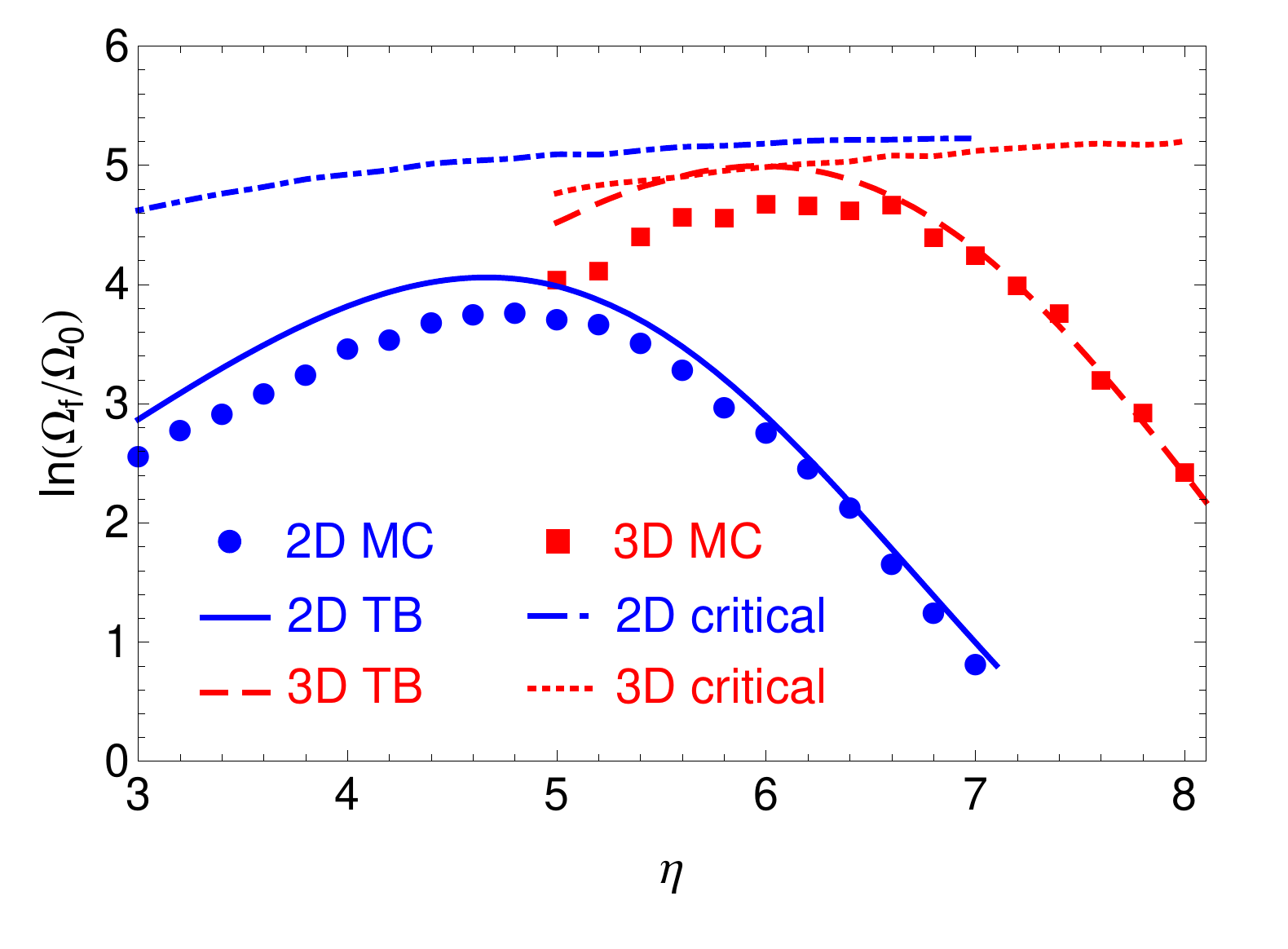} 
  \caption{\label{fig:pd2d3d}(Color online) Comparison of the efficiency of
    evaporative cooling in two-dimensional (2D) and three-dimensional (3D) harmonic
    traps. $\text{ln}(\Omega_{\rm{f}}/\Omega_0)$'s are calculated at
    $N_{\rm{f}}/N_0=0.1$, and $\zeta=1/200$. Red (blue) line: 3D (2D)
    harmonic trap from method in Appendix \ref{sec:tb}; red squares:
    3D harmonic trap from MC simulation; blue disks: 2D harmonic trap
    from MC simulation; blue (red) dotted line: $\Omega_c$ for a 2D (3D)
    trap.}
\end{figure}

To incorporate such differences, we use MC simulation and a
truncated-Boltzmann (TB) method based on kinetic theory (see
Appendix~\ref{sec:tb}) to calculate the evaporation
trajectories. Here, we consider harmonic traps and isotropic
energy-independent collisions in order to isolate the effect of
dimensionality. Following a similar procedure to that of the previous
section, we calculate $\text{ln}(\Omega_{\rm{f}}/\Omega_0)$ at fixed
$N_{\rm{f}}/N_0$, under equivalent initial conditions for both types
of traps. As shown in FIG.~\ref{fig:pd2d3d}, both the MC and TB
approaches show a lower achievable increase in the phase-space density
for a two-dimensional harmonic trap. This suggests that evaporative
cooling is intrinsically less efficient in two-dimensional traps. The
discrepancy between MC and TB, which increases for smaller~$\eta$, is
caused by the discrepancy between the form of the energy
distributions, which for TB is constrained. At smaller $\eta$, the
truncated Boltzmann distribution deviates significantly from the actual
distribution of molecules, as calculated in MC.

\section{\label{sec:KRb}Evaporative cooling of KRb molecules}

In a quasi-two-dimensional trap, the elastic and reactive collisions
between KRb molecules both depend on both the strength of the external
electric field and the confinement induced by the lattice. Under
current trapping conditions in experiments, a favorable ratio between
the elastic and reactive processes can be reached at a moderate
electric field~\cite{stereoloss,goulvenband, goulven2d}. We show in
FIG.~\ref{fig:kappa}(b) the total scattering cross-section for the
elastic process and reactive process of KRb molecules at $d=0.2$~debye
with the confinement along the lattice direction given by
$\nu=23$~kHz. In the ultracold regime, the elastic cross-section
scales as $E_{\rm{c}}^{3/2}$, while the reactive cross-section scales
as $E_{\rm{c}}^{1/2}$~\cite{li2008}. Furthermore, elastic processes
are found to be generally faster than reactive processes, supporting
 the potential for successful evaporative cooling~\cite{goulven2d} while the large elastic
cross-section favors fast rethermalization.
\begin{figure}\centering
  \includegraphics[width=0.31\textwidth]{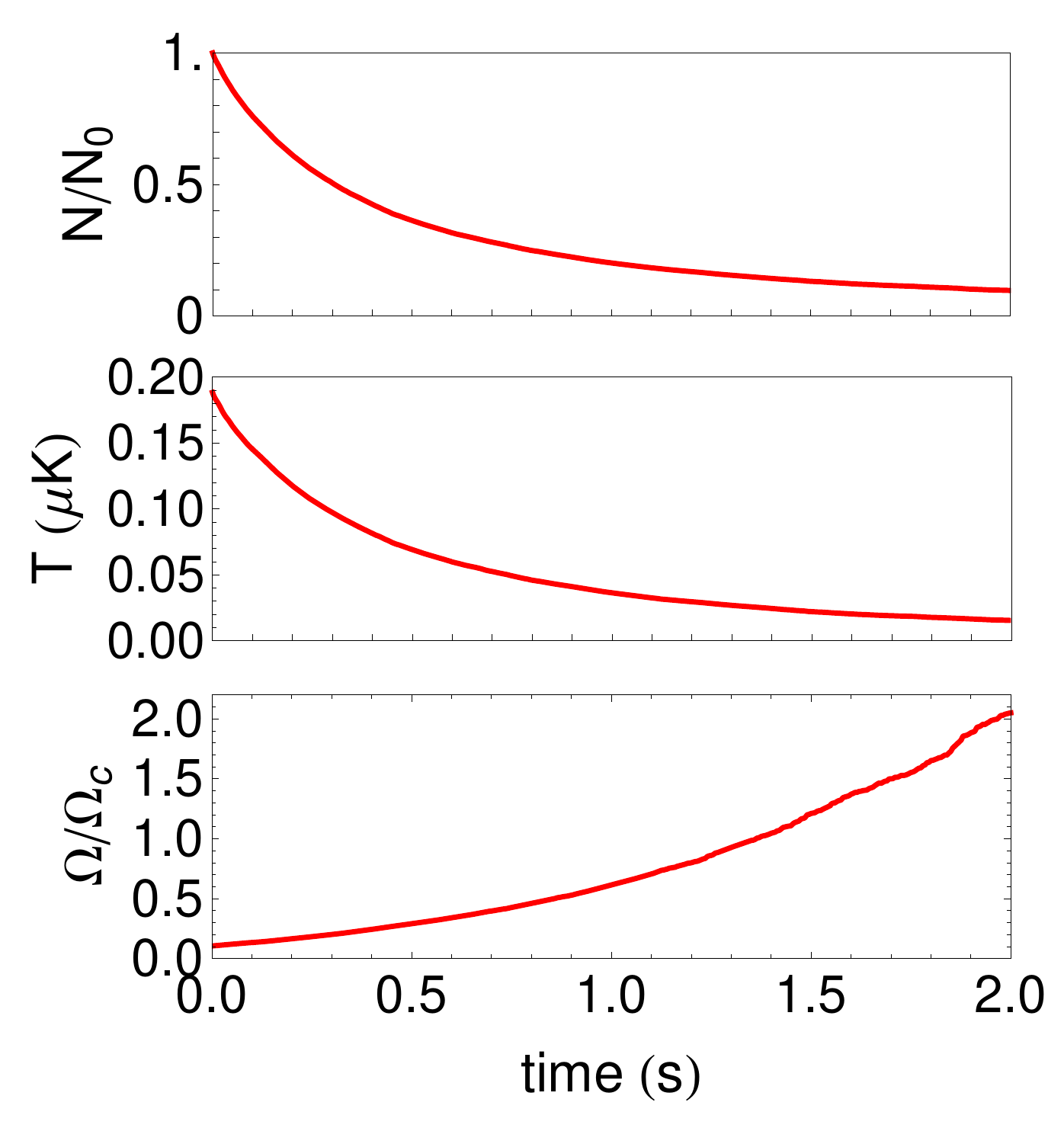} \caption{\label{fig:pkrb}(Color
    online) Evaporative cooling trajectories for KRb molecules inside
    a 2D trap of $2\pi\times20$~Hz, with scattering cross-sections as
    computed in FIG.~\ref{fig:kappa}(b). The initial temperature
    was~$\sim200$~nK, with an initial phase-space density
    $\Omega_0\sim0.1\Omega_c$, and $\eta=3.8$.}.

\end{figure}
The quantitative knowledge of the elastic and reactive collisions
between KRb molecules allows us to apply MC simulation with realistic
experimental parameters. In FIG.~\ref{fig:pkrb}, we show evaporation
trajectories in a $2\pi\times 20$~Hz two-dimensional gaussian
trap. For this simulation, we have assumed the instantaneous removal
of energetic molecules above the cut, as in
Sec.~\ref{subsec:anisoevap} and Sec.~\ref{sec:2dvs3d}. We also assume
 reactive losses as discussed above. The initial temperature was chosen to be within
the accessible regime of experiments. The results show that a
considerable increase in phase-space density can be achieved via
evaporative cooling. We note that there are further experimental
effects that can in principle reduce the achieved phase-space density
increase from that calculated here, such as the finite rate of
removing energetic molecules and other sources of loss and heating. On
the other hand, progress towards realizing deeper lattices ({\em
  i.e.}, $\nu>23$~kHz), can potentially reduce~$\zeta$ and thereby
enhance the cooling efficiency.

\section{\label{sec:antievap}Anti-evaporation in quasi-2D}

\begin{figure}\centering
\includegraphics[width=0.49\textwidth]{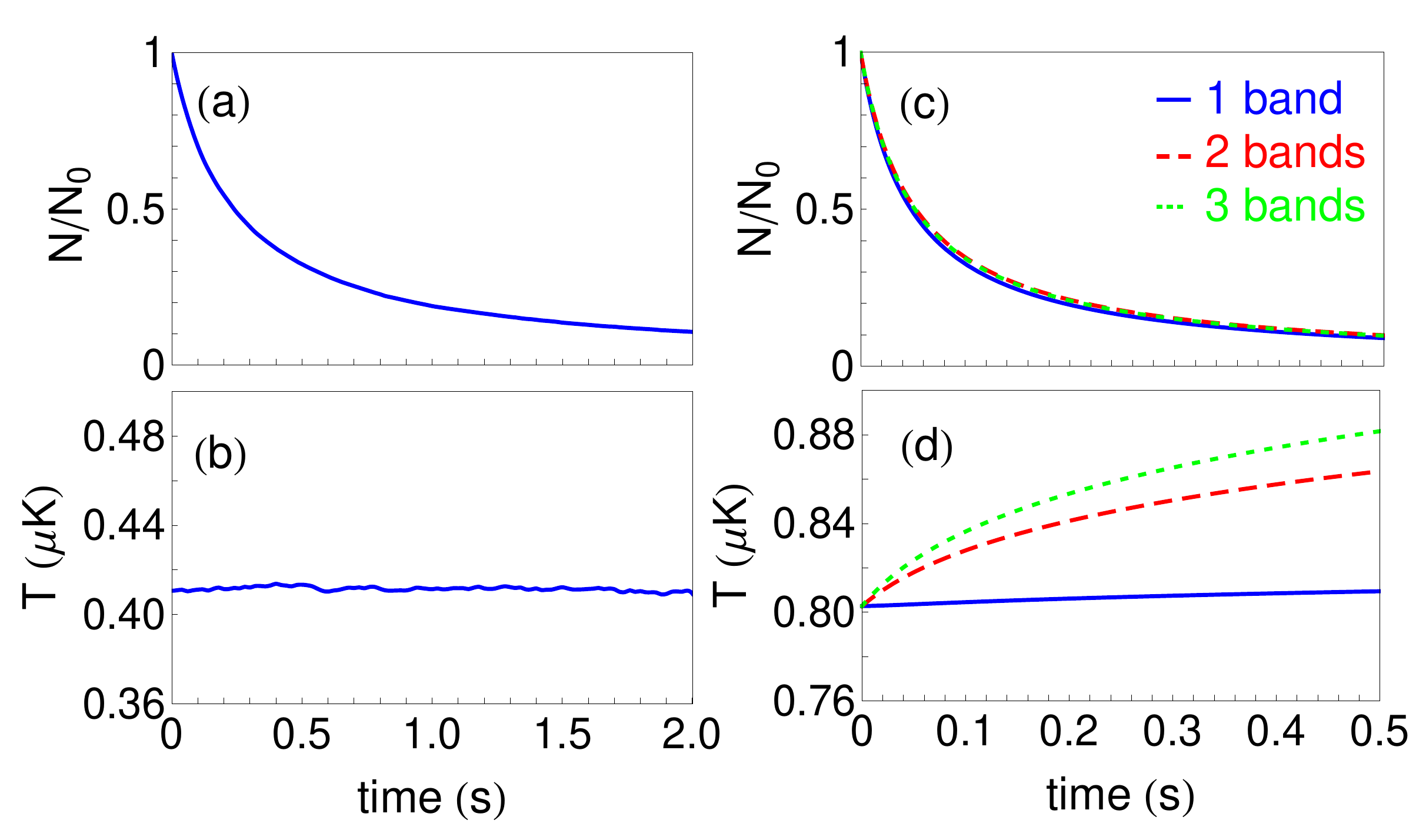}
\caption{\label{fig:antievap}(Color online) Heating due to two-body
  losses in a two-dimensional harmonic trap. Here (a) and (b) show the
  MC result in a two-dimensional harmonic trap where all molecules are
  in the lowest band of the lattice potential. The results for (c) and
  (d) are obtained with the master equation approach. The blue line
  assumes only 1 band along the lattice direction is populated. The
  red dashed line (green dots) assumes 2 (3) bands are populated. $N_0$
  is the initial total molecule number. The initial temperature  is chosen to be around 800~nK, so that when two bands are included, there are $\sim$25\% of molecules on the second band, and when three bands are included, there are $\sim$19\% of molecules on the second band and $\sim$6\% on the third band.}.
\end{figure}
\begin{figure}\centering
\includegraphics[width=0.36\textwidth]{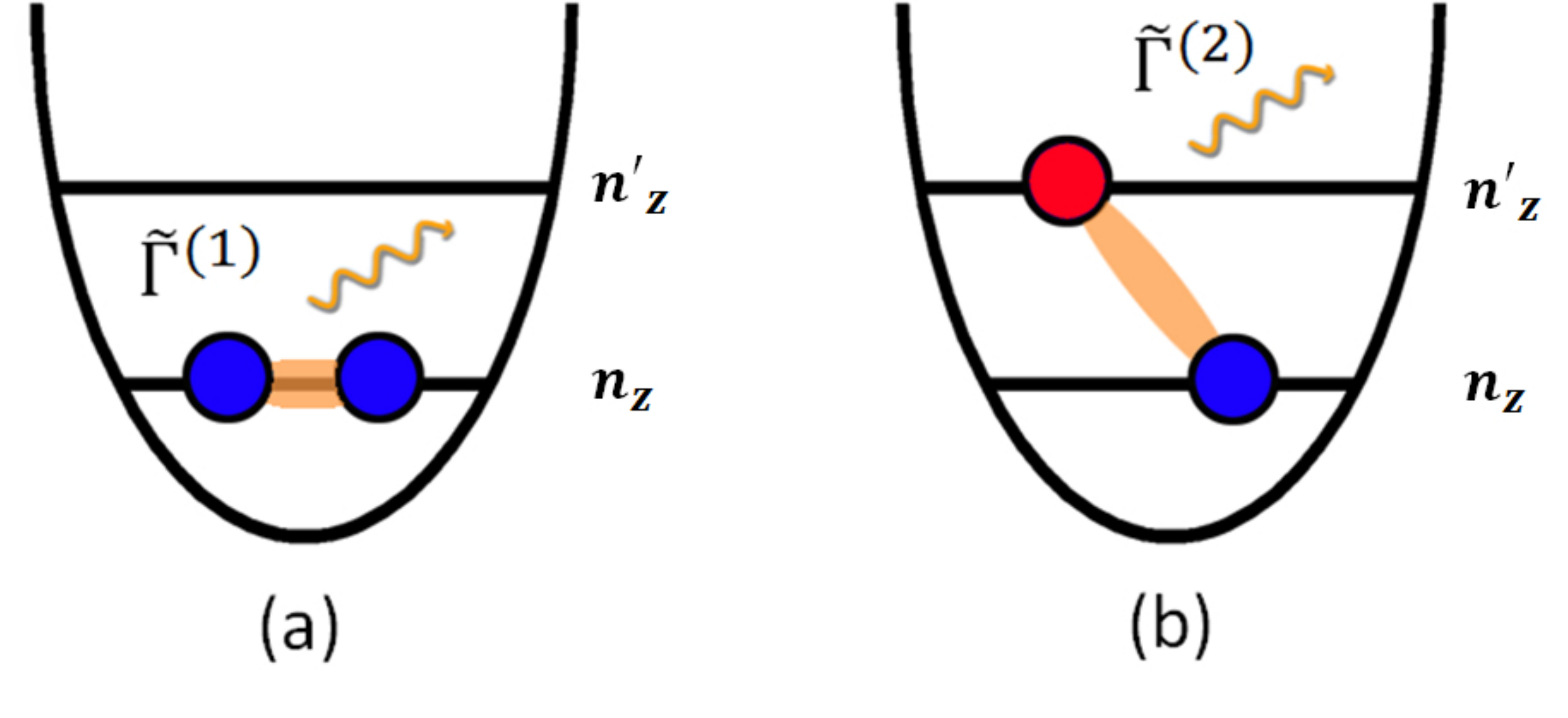}
\caption{\label{fig:colband}(Color online) Collisions between
  molecules in a quasi-2D trap. (a) Intra-band collisions, occuring at
  rate $\tilde{\Gamma}^{(1)}$.  (b) Inter-band collisions, occuring at
  rate $\tilde{\Gamma}^{(2)}$ (see Appendix~\ref{ap:master}). The wavy arrows indicate that the molecules are lost from the trap after these reactive collisions. }.

\end{figure}
Fast reactive collisions are disadvantageous for evaporative cooling,
since they not only reduce the molecule number, but can also cause
heating known as ``anti-evaporation"~\cite{ninature}. Since the ability
to identify the relative importance of different heating mechanisms is
important in the design of experiments, in this section we
quantitatively investigate heating due to two-body reactive
collisions. 

Anti-evaporation arises from the fact that reactive collisions are
more frequent in the high density region of the trap
(density-selection). Molecules there have on average total energies
that are less than the ensemble average. Thus one may anticipate
evaporative heating rather than evaporative cooling to occur. In
competition with this, however, is the effect that the reactive
collision cross-section is dependent on the collision energy, {\em i.e.}, molecules with energies higher than the average are more
likely to engage in a reactive collision and be removed
(velocity-selection). In three-dimensional harmonic traps, the reactive processes for ultracold indistinguishable fermionic KRb molecules are mainly $p$-wave two-body collisions, the cross-section of which scales as $E_{\rm{c}}^{1/2}$~\cite{ninature}. The density-selection mechanism wins over the velocity-selection mechanism and consequently there is net heating induced by losses~\cite{ninature}. In two-dimensional trapping potentials, however, this is not necessarily true even though the cross-section scales similarly with $E_{\rm{c}}$ (see Sec.~\ref{sec:KRb}), as we now show.

We consider a two-dimensional harmonic trap and assume a Boltzmann
distribution as given in Eq.~(\ref{eq:bolt}), for which the number
loss rate is
\begin{equation}
\frac{dN}{dt}=-2\int d^2xd^2p_1d^2p_2\lambda(E_{\rm{rel}})v_{\rm{rel}}f_0(\mathbf{x},\mathbf{p}_1)f_0(\mathbf{x},\mathbf{p}_2),\label{eq:dn}
\end{equation}
and the corresponding energy loss is
\begin{equation}
\frac{dE}{dt}\!=\!-\!\!\!\int\!\!d^2xd^2p_1\!d^2p_2\lambda(E_{\rm{rel}})v_{\rm{rel}}f_0(\mathbf{x},\!\mathbf{p}_1)f_0(\mathbf{x},\!\mathbf{p}_2)(E_1\!+\!E_2),\label{eq:de}
\end{equation}
where $E_{\rm{rel}}$ and $v_{\rm{rel}}$ are the relative energy and
velocity between the two coliding molecules, and $E_1+E_2$ is the
total energy. From Eq.~(\ref{eq:dn}) and Eq.~(\ref{eq:de}), with
$\lambda(E_{\rm{rel}})\propto E^{1/2}$ (see Sec.~\ref{sec:KRb}), one
can derive the energy loss per molecule, which is given by
\begin{equation}
\frac{dE}{dN}=2k_BT.\label{eq:dedn}
\end{equation}
This implies that there is no net heating or cooling in a pure
two-dimensional harmonic trap due to reactive loss. This is because
the energy loss in a single reactive process is exactly the average
energy for a pair of molecules. In contrast, a similar calculation for
three-dimensional harmonic traps gives $dE/dN=2.75k_BT$, which is
smaller than the average energy per molecule ({\em i.e.} $3k_BT$),
indicating heating in that case. This conclusion for three-dimensional
traps is consistent with the experimental observation in
Ref.~\cite{ninature} (heating rate $\sim$100~nK/s).

However, we emphasize that Eq.~(\ref{eq:dn}) to Eq.~(\ref{eq:dedn})
only takes into account the two-body reactive collisions and assumes a
Boltzmann distribution, which implies sufficiently fast
rethermalization. Note that at zero electric field, the elastic
collisions that lead to thermalization are infrequent and can
generally be neglected in experiments with KRb
molecules~\cite{ninature}, thus a Boltzmann distribution is not
guaranteed. 

To study possible effects caused by the deviation from a Boltzmann
distribution during the evolution, we use MC simulations to calculate
the molecule loss and temperature change in a two-dimensional harmonic trap, assuming
only two-body reactive processes between molecules. The results are
plotted in FIGs.~\ref{fig:antievap}(a) and (b) respectively. The reactive collision
rate is taken from FIG.~\ref{fig:kappa} (b). The heating rate is found
to be very small, illustrating a drastically different quantitative
behavior from that seen in three-dimensional traps.

The periodic array of two-dimensional pancake traps for confining KRb
molecules is generated by a one-dimensional optical
lattice~\cite{stereoloss}. This forms energy bands separated by
$\sim\hbar\omega_z$, where $\omega_z=2\pi\nu$ and $\nu$ is the frequency of the harmonic approximation of the lattice potential. When the
temperature is sufficiently low, the longitudinal degree-of-freedom
along the lattice axis is effectively frozen out and one may consider
solely the lowest band. However, as the temperature increases, higher
bands can be populated. For example, at $T\sim800$~nK in a lattice of
$\nu=23$~kHz, about 25\% of molecules occupy higher bands.

With multiband population, collisions between molecules can occur as
inter-band or intra-band, as shown in FIG.~\ref{fig:colband}.  The
inter-band collisions involve additional degrees-of-freedom along the
longitudinal direction (see FIG.~\ref{fig:colband}(b)), and thus do
not satisfy the assumptions leading to Eq.~(\ref{eq:dedn}) and can
give rise to anti-evaporation heating~\cite{ninature}.

In the ultra-low temperature regime, these collisions are dominated by the lowest odd partial wave~\cite{ninature}. We adopt a $p$-wave many-body Hamiltonian to describe such reactive interactions~\cite{pwaveH,anascience2011,sr2013}:
\begin{eqnarray}
  \hat{H}_{p}=&&\frac{3\pi b_p^3\hbar^2}{m}\int d^3r
  [(\nabla\hat{\Psi}^\dagger(\mathbf{r}))\hat{\Psi}^\dagger(\mathbf{r})-\hat{\Psi}^\dagger(\mathbf{r})
  (\nabla\hat{\Psi}^\dagger(\mathbf{r}))]\nonumber\\
  &&\cdot
  [(\nabla\hat{\Psi}(\mathbf{r}))\hat{\Psi}(\mathbf{r})-\hat{\Psi}(\mathbf{r})(\nabla\hat{\Psi}(\mathbf{r}))],
  \label{eq:pwaveH}
\end{eqnarray}
where $\hat{\Psi}(\mathbf{r})$ is a fermionic field operator that annihilates a molecule at position $\mathbf{r}$, and $b_p^3$ is the $p$-wave inelastic scattering volume. The value of $b_p$ for KRb
molecules can be found from the reactive cross-section calculated in
the previous section. In a trap with transverse confinement $\omega_r$
and longitudinal confinement $\omega_z$, the field operator can be
expanded in the basis of non-interacting harmonic oscillator
eigenstates $\psi^r_n$ (corresponding to $\omega_r$) and $\psi^z_n$ (corresponding to $\omega_z$) as:
\begin{equation}
  \hat{\Psi}(\mathbf{r})=\sum_{\mathbf{n}}\hat{c}_{\mathbf{n}}
  \psi_{n_x}^r(x)\psi^r_{n_y}(y)\psi^z_{n_z}(z),
\end{equation}
where  $\mathbf{n}=(n_x,n_y,n_z)$ enumerates the mode number along each
dimension, and $\hat{c}_{\mathbf{n}}$  annihilates a fermion in mode $\mathbf{n}$. 

Considering elastic collisions to be negligible, we assume there is no
change of modes during the losses, {\em i.e.}, a molecule initially in
mode $\mathbf{n}$ remains in mode $\mathbf{n}$. With such an
assumption, each molecule can be labeled with its mode indices, and
for identical fermions, there are no two molecules with the same
indices. The two-body losses can then be accounted for by jump
operators
$\hat{A}_{\mathbf{n},\mathbf{m}}=\sqrt{\Gamma_{\mathbf{n},\mathbf{m}}}
\hat{c}_{\mathbf{n}}\hat{c}_{\mathbf{m}}$, which remove two molecules
in mode $\mathbf{n}$ and $\mathbf{m}$ at a rate
$\Gamma_{\mathbf{n},\mathbf{m}}$ determined by the corresponding $p$-wave two-body reactive collisions Eq.~(\ref{eq:pwaveH}) (see Appendix~\ref{ap:master} for
details). When $n_z=m_z$, these correspond to the intra-band collisions
(FIG.~\ref{fig:colband}(a)); when $n_z\neq m_z$, these correspond to
the inter-band collisions (FIG.~\ref{fig:colband}(b)).

The dynamics can then be described  by the quantum master equation
\begin{equation}
  \frac{d\hat{\rho}}{dt}=\frac{1}{2}\sum_{\mathbf{n},\mathbf{m}}
  (2\hat{A}_{\mathbf{n},\mathbf{m}}\hat{\rho}\hat{A}^\dagger_{\mathbf{n},\mathbf{m}}
  -\hat{A}^\dagger_{\mathbf{n},\mathbf{m}}\hat{A}_{\mathbf{n},\mathbf{m}}\hat{\rho}
  -\hat{\rho}\hat{A}^\dagger_{\mathbf{n},\mathbf{m}}\hat{A}_{\mathbf{n},\mathbf{m}})
  .\label{eq:master}
\end{equation}
Since we have assumed that the molecules do not change modes during
evolution, in order to solve Eq.(~\ref{eq:master}), we can use
$\hat{\rho}=\prod_i\sum_{\mathbf{n}}\rho^i_{\mathbf{n},\mathbf{n}}\delta
(\mathbf{n}-\mathbf{n}^i)|\mathbf{n}\rangle\langle \mathbf{n}|$, where
$i=1,2,3...N$ represents the $i$th molecule that initially is in mode
$\mathbf{n}^i$, $N$ is the initial total number of molecules. Since
the quantum jumps here correspond to the reactive loss of pairs of
molecules, the off-diagonal coherence terms have been dropped. The
relevant equation of motion under these assumptions is
\begin{eqnarray}
  \frac{d\rho^i_{\mathbf{n}^i,\mathbf{n}^i}}{dt}=
  -\rho^i_{\mathbf{n}^i,\mathbf{n}^i}&&\sum_{j\neq i}
  \Gamma_{\mathbf{n}^i,\mathbf{m}^{j}}\rho^j_{\mathbf{m}^j,\mathbf{m}^j}.
\label{eq:motion}
\end{eqnarray}
The total molecule number at time $t$ is thus
\begin{equation}
N(t)=\sum_i\rho^i_{\mathbf{n}^i,\mathbf{n}^i}(t),
\end{equation}
and the average temperature  can be found from
\begin{equation}
  T(t)=\frac{1}{N(t)k_B}\sum_i\hbar\omega_r(n^i_x+n^i_y)
  \rho^i_{\mathbf{n}^i,\mathbf{n}^i}(t).
\end{equation}
Furthermore, the initial population in different bands can be
determined by
\begin{eqnarray}
\frac{N^\alpha}{N}=\frac{e^{-(1/2+\alpha\hbar\omega_z)/k_BT}}
{\sum_\alpha e^{- (1/2+\alpha\hbar\omega_z)/k_BT}},\label{eq:band}
\end{eqnarray}
where $\alpha=0,1,...$ is the band index. For the initial condition,
the density matrix is simulated such that $N^\alpha$ molecules are
assigned to band index $n^i_z=\alpha$, and the transverse modes
$n^i_x, n^i_y$ are randomly generated from a Boltzmann distribution
with given initial temperature $T$. The final result is averaged over
many different simulations. In order to show the effect of including different number of bands, we assum a cut-off in the highest band index. In our calculation, if a total of $n_{max}$ bands are taken into account, the population of the lower $n_{max}-2$ bands is determined according to Eq.~(\ref{eq:band}). All of the reminder molecules are assigned to the highest populated band, $n_{max}-1$. 
     
In FIG.~\ref{fig:antievap}(c) and (d), we show the result with
temperature $T=800$~nK and lattice frequency $\nu=23$~kHz, which are
typical for KRb experiments. While the small amount of multiband
population does not lead to a significant change in the loss rate for
molecules (FIG.~\ref{fig:antievap}(c)), there can nevertheless be
significant heating. In consequence, it is necessary to greatly
suppress the population outside the lowest band for evaporative
cooling experiments to be effectively performed in
quasi-two-dimensional geometries when reactive loss of the form
considered here is present.

\section{\label{sec:conclusion}Conclusion}
Motivated by recent experiments of KRb molecules, we have applied MC
simulation methods and semianalytical approaches to perform a detailed
study of evaporative cooling properties in two-dimensional traps with
anisotropic collisions. We have quantitatively analyzed the dependence
of the thermalization rate on the anisotropy of the elastic
collisions. Specifically, for our calculations of the differential and
total scattering cross-section for KRb molecules confined in a
quasi-two-dimensional trap, we were able to investigate the efficiency
of evaporative cooling for the practical parameter regime of recent
experiments. The dipole-dipole interactions resulted in highly
anisotropic elastic collisions which were disadvantageous for
evaporative cooling when compared with isotropic collisions. 

The reduced dimension of the trapping potential further decreased the
efficiency of evaporative cooling when compared with conventional
evaporation procedures for three-dimensional traps. Nevertheless, we
showed that the phase space density of KRb molecules can potentially
be increased using evaporative cooling with parameters that are
accessible under current experimental conditions. Although this could
be limited by further complicated real-world experimental details that
we have not considered, it is also the case that future experimental
progress in increasing the ratio between the elastic and reactive
collision rates could lead to more efficient evaporative cooling of
KRb molecules.  

We also developed theoretical models to investigate the
anti-evaporation induced by two-body losses. Our results highlighted
the distinctions between the evaporation processes in two-dimensional
and three-dimensional geometries through the role of multiband
excitations in a quasi-two-dimensional trap. We point out that the
importance of multiband physics has also been addressed in recent
literature~\cite{soltan2011,mark2011,will2010}.
\begin{acknowledgments}
  We would like to thank the JILA KRb
  group, John Bohn, Kaden Hazzard and Alexander Pikovski for helpful
  discussions. We acknowledge support from ARO, AFOSR, NSF-PIF, NSF JILA-PFC-1125844, NIST, and ARO-DARPA-OLE.
\end{acknowledgments}

\appendix
\section{\label{append:cross}Scattering cross-section in a quasi-2D
  trap}

Here we outline the theoretical formalism we used to compute the total
and differential scattering cross-sections in a two-dimensional space,
for elastic and reactive processes of KRb+KRb collisions, in the
presence of an electric field along the confinement direction. More
details can also be found in Ref.~\cite{goulvenband}.

We use a time-independent scattering formalism based on Jacobi
coordinates. The KRb molecules are assumed to be in their ground
electronic, vibrational and rotational states. They are also assumed
to be in the ground state of the one dimensional optical lattice. The
results of this formalism compare well with experimental
observation~\cite{stereoloss}. The vector $\mathbf{r}$ represents the
relative distance between the two centers of mass of the two
molecules. The potential energy between the two molecules is given by
the van der Waals interaction and the dipole-dipole interaction that
arises in the presence of an external applied electric field. In
addition to the intermolecule interaction potential, each molecule
feels a one dimensional confining potential approximated by an
harmonic oscillator. At short range, we apply an absorbing potential
that takes into account the fact that the KRb molecules are chemically
reactive and are lost from the trap. 

The time-independent Schr\"{o}dinger equation is solved for each
intermolecular separation $r$ using a spherical representation of
$\mathbf{r}=(r,\theta,\phi)$. At long distance, the van der Waals and
dipolar interactions vanish while the confinement is still present. It
is therefore more convenient to use a cylindrical representation of
$\mathbf{r}=(\rho,\phi,z)$. Since the electric field and the
confinement axis of the optical lattice are parallel to the
quantization axis, the quantum number $m_l$ associated with the
projection of the orbital angular momentum is strictly conserved
during the collision. Since we assume that the fermionic KRb
molecules, all in the same internal ground state, are also in the same
external ground state of the optical lattice, the quantum number $m_l$
is restricted to odd values $m_l=\pm 1,\pm 3,\pm 5,
...$~\cite{goulvenband}. A frame transformation is performed at long
range between the two representations. Applying asymptotic boundary
conditions at long range, we obtain the scattering matrix $S^{m_l}$ and
the transition matrix $T^{m_l}=S^{m_l}-1$.

The elastic differential cross-section for initial and final KRb
molecules in the internal and external ground state is given
by~\cite{lapidus1982}:
\begin{equation}
\frac{d\lambda(\phi)}{d\phi}=|f_s(\phi)|^2
\end{equation}
where the scattering amplitude is:
\begin{equation}
  f_s(\phi)=\left(\frac{1}{2\pi ik}\right)^{1/2}\sum_{m_l=1,3,...}^{\infty}
  \epsilon_{m_l} \text{cos}( m_l \phi)T^{m_l}\Delta
\end{equation}
with $k=\sqrt{2\mu E_{\rm{c}}}/\hbar$ denoting the wavevector,
$E_{\rm{c}}$ the collision energy, $\mu$ the reduced mass of the
colliding pair, $\epsilon_{m_l}=2$ if $m_l\ge 1$, $\epsilon=1$ if $m_l=0$, and
$\Delta=2$ or $1$ if the molecules are indistinguishable or not. To
converge the results, we used 10 odd values of $m_l$, with
$m_l=[1,3,\ldots,19]$. The total elastic cross-section is given by:
\begin{equation}
\lambda=\int_0^{2\pi}d\phi\frac{d\lambda(\phi)}{d\phi}.
\end{equation}
As seen from the differential cross-section shown in
FIG.~\ref{fig:kappa}(a), scattering in the perpendicular direction is
forbidden. This is due to the cos$(m_l\phi)$ term, with the odd $m_l$
restriction arising from the fermionic indistinguishable character of
the molecules.

\section{\label{sec:MC}Monte Carlo simulation}

With its stochastic nature, the Monte Carlo method is capable of
simulating the individual collisions that are intrinsic to evaporative
cooling phenomena~\cite{huang1996, pwaveMC, huang1997}, and this leads
to a flexible algorithm in which it is easy to incorporate a variety
of different conditions in a straightforward way. The detailed
simulation algorithm consists mainly of the following steps:

1. Preparation of an ensemble of particles with coordinates and
velocities generated from a given probability distribution, e.g. an
initial equilibrium Boltzmann distribution.

2. Evolution of the particles between collisions that follows the
classical Hamilton's equations of motion, {\em i.e.}
\begin{eqnarray}
\frac{d\mathbf{x}}{dt}&&=\mathbf{v},\nonumber\\
\frac{d\mathbf{v}}{dt}&&=-\nabla V(\mathbf{x}),
\end{eqnarray}
where $V(\mathbf{x})$ is the external potential. The time step is
chosen to be both small enough to guarantee numerical convergence for
the computed trajectory, as well as much less than the time between
adjacent collision events.

3. Checking every pair of particles to decide if there is a
collision. Since we are interested in two-body collisions, the
collision events are determined by the distance between
particles. This distance is characterized by the scattering
length~$a_s$. In three-dimensional collisions, there is a collision
cross-section with units of area, $\sigma=\pi a_{s}^2$. In
two-dimensional collisions, there is an analogous cross-section with
units of length, $\lambda=2a_s$.

4. Changing the state of the pair of particles if they collide. For an
elastic collision event, the velocities after collisions are
determined from the conservation of total momentum and energy. The
scattering angle is simulated from a probability distribution
determined by the differential cross-section. For instance, in
two-dimensional $s$-wave scattering, the outgoing angles $\phi$ are
uniformly distributed in the interval $[0,2\pi)$, while for $p$-wave
scattering, the distribution of the outgoing angles follows the
distribution $\text{cos}^2\phi$. In three-dimensional isotropic
scattering, there are two random angles: an azimuthal angle $\phi$
uniformly distributed in the interval $[0,2\pi)$ and a polar angle
$\theta$ from uniformly distributed according to
$\text{cos}\theta$. The non-uniform distribution of the scattering
angles does not affect the collision rate that is determined by the
total cross-section, but the rate of redistributing energies does
vary. For a loss event corresponding to two-body
reactive collision, the two particles are lost from the trap after
the collision, and thus deleted from the simulation when such
processes occur.

5. Averaging over many initial samples and trajectories. The physical
quantities such as the total number of particles, temperature,
collision rate and phase space density can be computed statistically
from the simulation ensemble.

\section{\label{sec:tb}Kinetics of evaporative cooling in 2D}

An alternative algorithm which is partially analytic can be derived by
assuming that in the process of evaporative cooling, the energetic
particles can be efficiently removed from the trap. The system is
assumed to follow a truncated Boltzmann distribution 
$f(\mathbf{x},\mathbf{p})$ (Eq.~\ref{eq:tbf}), with a cut-off energy
$\epsilon_t$ \cite{walraven, tbfermi}, meaning that there is no
particle with energy $E>\epsilon_t$. Then similar to the description
of an equilibrium ensemble, the measured quantities of the system can
be expressed by averaging over this distribution function. For
example, the collision rate is
\begin{eqnarray}
  \gamma&=&\frac{\lambda\Lambda^2}{m}\int_{\epsilon_1,\epsilon_2} 
  d^2xd^2p_1d^2p_2\nonumber\\
  &&{}\times f(\mathbf{x},\mathbf{p}_1)
  f(\mathbf{x},\mathbf{p}_2)|\mathbf{p}_1-\mathbf{p}_2|,\label{eq:c1}
\end{eqnarray}
in which $\Lambda=1/(2\pi\hbar)^2$, and
$\epsilon_{1,2}=p_{1,2}^2/2m+V(\mathbf{x}_{1,2})$ are the energies of
the incident particles. For simplicity, here we have assumed isotropic
energy-independent elastic and reactive collisions, so $\lambda$ in
Eq.~(\ref{eq:c1}) is constant. We also assume the two-body reactive
collisions happen at rate $\zeta\gamma$. During forced evaporative
cooling, $\epsilon_t$ decreases with time, so does the temperature
$T$, and the distribution function
$f=f(\mathbf{x},\mathbf{p})$. As a result, the evolution of the
system from $t$ to $t'$ can be modeled via three steps: the change due
to $f$ when $\epsilon_t$ decreases to $\epsilon'_{t}$ ($dN_1, dE_1$), the change due to evaporation ($dN_2,
dE_2$), and the change due to the reactive losses ($dN_3,
dE_3$). These can be represented by the following equations:
\begin{eqnarray}
dN_1=&&\Lambda\int_{\epsilon=\epsilon_t'}^{\epsilon_t}d^2xd^2pf,\\
dE_1=&&\Lambda\int_{\epsilon=\epsilon_t'}^{\epsilon_t}d^2xd^2p\epsilon f,\\
\frac{dN_{2}}{dt}=&&\frac{\lambda\Lambda^2}{m}\int_{\Sigma}  d^2xd^2p_1d^2p_2d\phi ' f_1f_2|\mathbf{p}_1-\mathbf{p}_2|,\\
\frac{dE_{2}}{dt}=&&\frac{\lambda\Lambda^2}{m}\int_{\Sigma}  d^2xd^2p_1d^2p_2d\phi 's f_1f_2|\mathbf{p}_1-\mathbf{p}_2|\epsilon_4,\\
\frac{dN_{3}}{dt}=&&\zeta \frac{\lambda\Lambda^2}{m}\int_{\Sigma'}  d^2xd^2p_1d^2p_2 f_1f_2|\mathbf{p}_1-\mathbf{p}_2|,\\
\frac{dE_{3}}{dt}=&&\zeta\frac{\lambda\Lambda^2}{m}\int_{\Sigma'} d^2xd^2p_1d^2p_2f_1f_2|\mathbf{p}_1-\mathbf{p}_2|(\epsilon_1 +\epsilon_2),\nonumber\\&&
\end{eqnarray}
where $f_1,f_2$ are the distribution function for each of the two
colliding particles, $\phi'$ specifies the scattering angle,
$\epsilon= \epsilon(\mathbf{x},\mathbf{p})$ is the energy of a
particle and is a function of the coordinate and momentum of the
particles, $\epsilon_3$ and $\epsilon_4$ are energies after collision,
which are determined once $\mathbf{p}_1$, $\mathbf{p}_2$ and $\phi'$
are known, and $\Sigma=\{\epsilon_1,\epsilon_2,\epsilon_3<\epsilon_t',
\epsilon_4>\epsilon_t'\}$ and
$\Sigma'=\{\epsilon_1,\epsilon_2<\epsilon_t'\}$ specify the
integration region.  These contributions add up to give the total
changes: $N(t')=N(t)-dN_1-dN_2-dN_3$ and
$E(t')=E(t)-dE_1-dE_2-dE_3$. The new temperature $T(t')$ is found from
\begin{eqnarray}
  N(t')=&&\Lambda\int_0^{\epsilon_t'}d^2xd^2p 
  f(\mathbf{x},\mathbf{p}),\\
  E(t')=&&\Lambda\int_0^{\epsilon_t'}d^2xd^2p 
  f(\mathbf{x},\mathbf{p})\epsilon.
\end{eqnarray}
Then the trajectory of evaporative cooling in two-dimensional traps can be solved
from the above equations. Similar calculations also apply to three-dimensional
harmonic traps, and the results are equivalent to solving the rate
equations following the method developed by Walraven {\em et al.}, as long
as the truncated Boltzmann distribution is a good
approximation~\cite{walraven}.

\section{
  \label{ap:master} Master equation approach for losses in a quasi-2D
  trap}

Represented in the basis of harmonic oscillator eigenstates, the rate coefficients
$\Gamma_{\mathbf{n},\mathbf{m}}$ defined in the jump operators in
Sec.~\ref{sec:antievap} can be written explicitly for intra-band
collisions ($n_z=m_z$, FIG.~\ref{fig:colband}(a))~\cite{sr2013}
\begin{eqnarray}
  \Gamma_{\mathbf{n},\mathbf{m}}=&&\tilde{\Gamma}^{(1)}\nonumber\\
  =&&\frac{3\sqrt{2\pi}b_p^3\sqrt{\omega_r\omega_z}}{a_{\rm{rho}}^3}
  Is(n_z,m_z,n_z,m_z)\nonumber\\&&\times \sum_{\sigma\neq\sigma '} 
  Is(n_\sigma,m_\sigma,n_\sigma,m_\sigma)
  Ip(n_{\sigma '},m_{\sigma '},n_{\sigma '},m_{\sigma '}),\nonumber\\
\end{eqnarray}
and for inter-band collisions ($n_z\neq m_z$,
FIG.~\ref{fig:colband}(b))
\begin{eqnarray}
  \Gamma_{\mathbf{n},\mathbf{m}}=&&\tilde{\Gamma}^{(2)}\nonumber\\
  =&&\frac{3\sqrt{2\pi}b_p^3\sqrt{\omega_r\omega_z}}{a_{\rm{rho}}^3}\sum_{\sigma_1\neq\sigma_2\neq\sigma_3}
  \frac{\omega_{\sigma_2}}{\omega_{r}}
  Is(n_{\sigma_1},m_{\sigma_1},n_{\sigma_1},m_{\sigma_1})\nonumber\\
  &&\times Ip(n_{\sigma_2},m_{\sigma_2},n_{\sigma_2},m_{\sigma_2})
  Is(n_{\sigma_3},m_{\sigma_3},n_{\sigma_3},m_{\sigma_3}),\nonumber\\
\end{eqnarray}
where $\sigma,\sigma'=\{x,y\}$, $\sigma_1,\sigma_2,\sigma_3
=\{x,y,z\}$, $a_{\rm{rho}}=\sqrt{\hbar/m\omega_r}$,
\begin{eqnarray}
  Is(n,m,p,q)=&&\int du\frac{e^{-2u^2}H_n(u)H_m(u)H_p(u)H_q(u)}
  {\pi\sqrt{2^{n+m+p+q}n!m!p!q!}},\nonumber\\
  \\
  Ip(n,m,p,q)=&&\int du\frac{e^{-2u^2} }{\pi
    \sqrt{2^{n+m+p+q}n!m!p!q!}}\nonumber\\
  &&\times [\partial H_n(u)H_m(u)-H_n(u)\partial H_m(u)]\nonumber\\
  &&\times [\partial H_p(u)H_q(u)-H_p(u)\partial H_q(u)],\nonumber\\
\end{eqnarray}
and $H_n,H_m,H_p,H_q$ are Hermite polynomials.

\bibliographystyle{apsrev}
\bibliography{paper}

\begin{thebibliography}{37}
\expandafter\ifx\csname natexlab\endcsname\relax\def\natexlab#1{#1}\fi
\expandafter\ifx\csname bibnamefont\endcsname\relax
  \def\bibnamefont#1{#1}\fi
\expandafter\ifx\csname bibfnamefont\endcsname\relax
  \def\bibfnamefont#1{#1}\fi
\expandafter\ifx\csname citenamefont\endcsname\relax
  \def\citenamefont#1{#1}\fi
\expandafter\ifx\csname url\endcsname\relax
  \def\url#1{\texttt{#1}}\fi
\expandafter\ifx\csname urlprefix\endcsname\relax\def\urlprefix{URL }\fi
\providecommand{\bibinfo}[2]{#2}
\providecommand{\eprint}[2][]{\url{#2}}

\bibitem[{\citenamefont{Lahaye et~al.}(2009)\citenamefont{Lahaye, Menotti,
  Santos, Lewenstein, and Pfau}}]{lahayephysics}
\bibinfo{author}{\bibfnamefont{T.}~\bibnamefont{Lahaye}},
  \bibinfo{author}{\bibfnamefont{C.}~\bibnamefont{Menotti}},
  \bibinfo{author}{\bibfnamefont{L.}~\bibnamefont{Santos}},
  \bibinfo{author}{\bibfnamefont{M.}~\bibnamefont{Lewenstein}},
  \bibnamefont{and} \bibinfo{author}{\bibfnamefont{T.}~\bibnamefont{Pfau}},
  \bibinfo{journal}{Rep. Prog. Phys.} \textbf{\bibinfo{volume}{72}},
  \bibinfo{pages}{126401} (\bibinfo{year}{2009}).

\bibitem[{\citenamefont{Sandars}(1967)}]{precision1}
\bibinfo{author}{\bibfnamefont{P.~G.~H.} \bibnamefont{Sandars}},
  \bibinfo{journal}{Phys. Rev. Lett.} \textbf{\bibinfo{volume}{19}},
  \bibinfo{pages}{1396} (\bibinfo{year}{1967}).

\bibitem[{\citenamefont{Hudson et~al.}(2006)\citenamefont{Hudson, Lewandowski,
  Sawyer, and Ye}}]{precision2}
\bibinfo{author}{\bibfnamefont{E.~R.} \bibnamefont{Hudson}},
  \bibinfo{author}{\bibfnamefont{H.~J.} \bibnamefont{Lewandowski}},
  \bibinfo{author}{\bibfnamefont{B.~C.} \bibnamefont{Sawyer}},
  \bibnamefont{and} \bibinfo{author}{\bibfnamefont{J.}~\bibnamefont{Ye}},
  \bibinfo{journal}{Phys. Rev. Lett.} \textbf{\bibinfo{volume}{96}},
  \bibinfo{pages}{143004} (\bibinfo{year}{2006}).

\bibitem[{\citenamefont{Yelin et~al.}(2006)\citenamefont{Yelin, Kirby, and
  C\^ot\'e}}]{moleculeQM}
\bibinfo{author}{\bibfnamefont{S.~F.} \bibnamefont{Yelin}},
  \bibinfo{author}{\bibfnamefont{K.}~\bibnamefont{Kirby}}, \bibnamefont{and}
  \bibinfo{author}{\bibfnamefont{R.}~\bibnamefont{C\^ot\'e}},
  \bibinfo{journal}{Phys. Rev. A} \textbf{\bibinfo{volume}{74}},
  \bibinfo{pages}{050301} (\bibinfo{year}{2006}).

\bibitem[{\citenamefont{DeMille}(2002)}]{mQM}
\bibinfo{author}{\bibfnamefont{D.}~\bibnamefont{DeMille}},
  \bibinfo{journal}{Phys. Rev. Lett.} \textbf{\bibinfo{volume}{88}},
  \bibinfo{pages}{067901} (\bibinfo{year}{2002}).

\bibitem[{\citenamefont{Micheli et~al.}(2006)\citenamefont{Micheli, Brennen,
  and Zoller}}]{moleculetoolbox}
\bibinfo{author}{\bibfnamefont{A.}~\bibnamefont{Micheli}},
  \bibinfo{author}{\bibfnamefont{G.}~\bibnamefont{Brennen}}, \bibnamefont{and}
  \bibinfo{author}{\bibfnamefont{P.}~\bibnamefont{Zoller}},
  \bibinfo{journal}{Nature Physics} \textbf{\bibinfo{volume}{2}},
  \bibinfo{pages}{341} (\bibinfo{year}{2006}).

\bibitem[{\citenamefont{Carr et~al.}(2009)\citenamefont{Carr, DeMille, Krems,
  and Ye}}]{carrcold}
\bibinfo{author}{\bibfnamefont{L.}~\bibnamefont{Carr}},
  \bibinfo{author}{\bibfnamefont{D.}~\bibnamefont{DeMille}},
  \bibinfo{author}{\bibfnamefont{R.}~\bibnamefont{Krems}}, \bibnamefont{and}
  \bibinfo{author}{\bibfnamefont{J.}~\bibnamefont{Ye}}, \bibinfo{journal}{New
  Journal of Physics} \textbf{\bibinfo{volume}{11}}, \bibinfo{pages}{055049}
  (\bibinfo{year}{2009}).

\bibitem[{\citenamefont{Ni et~al.}(2008)\citenamefont{Ni, Ospelkaus,
  De~Miranda, Pe'er, Neyenhuis, Zirbel, Kotochigova, Julienne, Jin, and
  Ye}}]{ni2008}
\bibinfo{author}{\bibfnamefont{K.}~\bibnamefont{Ni}},
  \bibinfo{author}{\bibfnamefont{S.}~\bibnamefont{Ospelkaus}},
  \bibinfo{author}{\bibfnamefont{M.}~\bibnamefont{De~Miranda}},
  \bibinfo{author}{\bibfnamefont{A.}~\bibnamefont{Pe'er}},
  \bibinfo{author}{\bibfnamefont{B.}~\bibnamefont{Neyenhuis}},
  \bibinfo{author}{\bibfnamefont{J.}~\bibnamefont{Zirbel}},
  \bibinfo{author}{\bibfnamefont{S.}~\bibnamefont{Kotochigova}},
  \bibinfo{author}{\bibfnamefont{P.}~\bibnamefont{Julienne}},
  \bibinfo{author}{\bibfnamefont{D.}~\bibnamefont{Jin}}, \bibnamefont{and}
  \bibinfo{author}{\bibfnamefont{J.}~\bibnamefont{Ye}},
  \bibinfo{journal}{science} \textbf{\bibinfo{volume}{322}},
  \bibinfo{pages}{231} (\bibinfo{year}{2008}).

\bibitem[{\citenamefont{Ospelkaus et~al.}(2009)\citenamefont{Ospelkaus, Ni,
  De~Miranda, Neyenhuis, Wang, Kotochigova, Julienne, Jin, and
  Ye}}]{coldpolar2}
\bibinfo{author}{\bibfnamefont{S.}~\bibnamefont{Ospelkaus}},
  \bibinfo{author}{\bibfnamefont{K.}~\bibnamefont{Ni}},
  \bibinfo{author}{\bibfnamefont{M.}~\bibnamefont{De~Miranda}},
  \bibinfo{author}{\bibfnamefont{B.}~\bibnamefont{Neyenhuis}},
  \bibinfo{author}{\bibfnamefont{D.}~\bibnamefont{Wang}},
  \bibinfo{author}{\bibfnamefont{S.}~\bibnamefont{Kotochigova}},
  \bibinfo{author}{\bibfnamefont{P.}~\bibnamefont{Julienne}},
  \bibinfo{author}{\bibfnamefont{D.}~\bibnamefont{Jin}}, \bibnamefont{and}
  \bibinfo{author}{\bibfnamefont{J.}~\bibnamefont{Ye}},
  \bibinfo{journal}{Faraday Discussions} \textbf{\bibinfo{volume}{142}},
  \bibinfo{pages}{351} (\bibinfo{year}{2009}).

\bibitem[{\citenamefont{Ni et~al.}(2010)\citenamefont{Ni, Ospelkaus, Wang,
  Qu{\'e}m{\'e}ner, Neyenhuis, De~Miranda, Bohn, Ye, and Jin}}]{ninature}
\bibinfo{author}{\bibfnamefont{K.}~\bibnamefont{Ni}},
  \bibinfo{author}{\bibfnamefont{S.}~\bibnamefont{Ospelkaus}},
  \bibinfo{author}{\bibfnamefont{D.}~\bibnamefont{Wang}},
  \bibinfo{author}{\bibfnamefont{G.}~\bibnamefont{Qu{\'e}m{\'e}ner}},
  \bibinfo{author}{\bibfnamefont{B.}~\bibnamefont{Neyenhuis}},
  \bibinfo{author}{\bibfnamefont{M.}~\bibnamefont{De~Miranda}},
  \bibinfo{author}{\bibfnamefont{J.}~\bibnamefont{Bohn}},
  \bibinfo{author}{\bibfnamefont{J.}~\bibnamefont{Ye}}, \bibnamefont{and}
  \bibinfo{author}{\bibfnamefont{D.}~\bibnamefont{Jin}},
  \bibinfo{journal}{Nature} \textbf{\bibinfo{volume}{464}},
  \bibinfo{pages}{1324} (\bibinfo{year}{2010}).

\bibitem[{\citenamefont{Bradley et~al.}(1995)\citenamefont{Bradley, Sackett,
  Tollett, and Hulet}}]{evap1}
\bibinfo{author}{\bibfnamefont{C.~C.} \bibnamefont{Bradley}},
  \bibinfo{author}{\bibfnamefont{C.~A.} \bibnamefont{Sackett}},
  \bibinfo{author}{\bibfnamefont{J.~J.} \bibnamefont{Tollett}},
  \bibnamefont{and} \bibinfo{author}{\bibfnamefont{R.~G.} \bibnamefont{Hulet}},
  \bibinfo{journal}{Phys. Rev. Lett.} \textbf{\bibinfo{volume}{75}},
  \bibinfo{pages}{1687} (\bibinfo{year}{1995}).

\bibitem[{\citenamefont{Davis et~al.}(1995)\citenamefont{Davis, Mewes, Andrews,
  van Druten, Durfee, Kurn, and Ketterle}}]{evap2}
\bibinfo{author}{\bibfnamefont{K.~B.} \bibnamefont{Davis}},
  \bibinfo{author}{\bibfnamefont{M.~O.} \bibnamefont{Mewes}},
  \bibinfo{author}{\bibfnamefont{M.~R.} \bibnamefont{Andrews}},
  \bibinfo{author}{\bibfnamefont{N.~J.} \bibnamefont{van Druten}},
  \bibinfo{author}{\bibfnamefont{D.~S.} \bibnamefont{Durfee}},
  \bibinfo{author}{\bibfnamefont{D.~M.} \bibnamefont{Kurn}}, \bibnamefont{and}
  \bibinfo{author}{\bibfnamefont{W.}~\bibnamefont{Ketterle}},
  \bibinfo{journal}{Phys. Rev. Lett.} \textbf{\bibinfo{volume}{75}},
  \bibinfo{pages}{3969} (\bibinfo{year}{1995}).

\bibitem[{\citenamefont{DeMarco and Jin}(1999)}]{evap3}
\bibinfo{author}{\bibfnamefont{B.}~\bibnamefont{DeMarco}} \bibnamefont{and}
  \bibinfo{author}{\bibfnamefont{D.}~\bibnamefont{Jin}},
  \bibinfo{journal}{Science} \textbf{\bibinfo{volume}{285}},
  \bibinfo{pages}{1703} (\bibinfo{year}{1999}).

\bibitem[{\citenamefont{Qu\'em\'ener and
  Bohn}(2010{\natexlab{a}})}]{goulvenloss}
\bibinfo{author}{\bibfnamefont{G.}~\bibnamefont{Qu\'em\'ener}}
  \bibnamefont{and} \bibinfo{author}{\bibfnamefont{J.~L.} \bibnamefont{Bohn}},
  \bibinfo{journal}{Phys. Rev. A} \textbf{\bibinfo{volume}{81}},
  \bibinfo{pages}{022702} (\bibinfo{year}{2010}{\natexlab{a}}).

\bibitem[{\citenamefont{de~Miranda et~al.}(2011)\citenamefont{de~Miranda,
  Chotia, Neyenhuis, Wang, Qu{\'e}m{\'e}ner, Ospelkaus, Bohn, Ye, and
  Jin}}]{stereoloss}
\bibinfo{author}{\bibfnamefont{M.}~\bibnamefont{de~Miranda}},
  \bibinfo{author}{\bibfnamefont{A.}~\bibnamefont{Chotia}},
  \bibinfo{author}{\bibfnamefont{B.}~\bibnamefont{Neyenhuis}},
  \bibinfo{author}{\bibfnamefont{D.}~\bibnamefont{Wang}},
  \bibinfo{author}{\bibfnamefont{G.}~\bibnamefont{Qu{\'e}m{\'e}ner}},
  \bibinfo{author}{\bibfnamefont{S.}~\bibnamefont{Ospelkaus}},
  \bibinfo{author}{\bibfnamefont{J.}~\bibnamefont{Bohn}},
  \bibinfo{author}{\bibfnamefont{J.}~\bibnamefont{Ye}}, \bibnamefont{and}
  \bibinfo{author}{\bibfnamefont{D.}~\bibnamefont{Jin}},
  \bibinfo{journal}{Nature Physics} \textbf{\bibinfo{volume}{7}},
  \bibinfo{pages}{502} (\bibinfo{year}{2011}).

\bibitem[{\citenamefont{Qu\'em\'ener and Bohn}(2011)}]{goulvenband}
\bibinfo{author}{\bibfnamefont{G.}~\bibnamefont{Qu\'em\'ener}}
  \bibnamefont{and} \bibinfo{author}{\bibfnamefont{J.~L.} \bibnamefont{Bohn}},
  \bibinfo{journal}{Phys. Rev. A} \textbf{\bibinfo{volume}{83}},
  \bibinfo{pages}{012705} (\bibinfo{year}{2011}).

\bibitem[{\citenamefont{Qu\'em\'ener and Bohn}(2010{\natexlab{b}})}]{goulven2d}
\bibinfo{author}{\bibfnamefont{G.}~\bibnamefont{Qu\'em\'ener}}
  \bibnamefont{and} \bibinfo{author}{\bibfnamefont{J.~L.} \bibnamefont{Bohn}},
  \bibinfo{journal}{Phys. Rev. A} \textbf{\bibinfo{volume}{81}},
  \bibinfo{pages}{060701} (\bibinfo{year}{2010}{\natexlab{b}}).

\bibitem[{\citenamefont{Ketterle and Druten}(1996)}]{ketterlereview}
\bibinfo{author}{\bibfnamefont{W.}~\bibnamefont{Ketterle}} \bibnamefont{and}
  \bibinfo{author}{\bibfnamefont{N.}~\bibnamefont{Druten}},
  \bibinfo{journal}{Advances in atomic, molecular, and optical physics}
  \textbf{\bibinfo{volume}{37}}, \bibinfo{pages}{181} (\bibinfo{year}{1996}).

\bibitem[{\citenamefont{Gochitashvili et~al.}(2010)\citenamefont{Gochitashvili,
  Kezerashvili, and Lomsadze}}]{goulvendipole}
\bibinfo{author}{\bibfnamefont{M.~R.} \bibnamefont{Gochitashvili}},
  \bibinfo{author}{\bibfnamefont{R.~Y.} \bibnamefont{Kezerashvili}},
  \bibnamefont{and} \bibinfo{author}{\bibfnamefont{R.~A.}
  \bibnamefont{Lomsadze}}, \bibinfo{journal}{Phys. Rev. A}
  \textbf{\bibinfo{volume}{82}}, \bibinfo{pages}{022702}
  (\bibinfo{year}{2010}).

\bibitem[{fno()}]{fnote}
\emph{\bibinfo{title}{\textup{Note: The cross-section in two-dimensions has a
  dimension of length.}}}

\bibitem[{\citenamefont{Monroe et~al.}(1993)\citenamefont{Monroe, Cornell,
  Sackett, Myatt, and Wieman}}]{thermcross}
\bibinfo{author}{\bibfnamefont{C.~R.} \bibnamefont{Monroe}},
  \bibinfo{author}{\bibfnamefont{E.~A.} \bibnamefont{Cornell}},
  \bibinfo{author}{\bibfnamefont{C.~A.} \bibnamefont{Sackett}},
  \bibinfo{author}{\bibfnamefont{C.~J.} \bibnamefont{Myatt}}, \bibnamefont{and}
  \bibinfo{author}{\bibfnamefont{C.~E.} \bibnamefont{Wieman}},
  \bibinfo{journal}{Phys. Rev. Lett.} \textbf{\bibinfo{volume}{70}},
  \bibinfo{pages}{414} (\bibinfo{year}{1993}).

\bibitem[{\citenamefont{DeMarco et~al.}(1999)\citenamefont{DeMarco, Bohn,
  Burke, Holland, and Jin}}]{pwaveMC}
\bibinfo{author}{\bibfnamefont{B.}~\bibnamefont{DeMarco}},
  \bibinfo{author}{\bibfnamefont{J.~L.} \bibnamefont{Bohn}},
  \bibinfo{author}{\bibfnamefont{J.~P.} \bibnamefont{Burke}},
  \bibinfo{author}{\bibfnamefont{M.}~\bibnamefont{Holland}}, \bibnamefont{and}
  \bibinfo{author}{\bibfnamefont{D.~S.} \bibnamefont{Jin}},
  \bibinfo{journal}{Phys. Rev. Lett.} \textbf{\bibinfo{volume}{82}},
  \bibinfo{pages}{4208} (\bibinfo{year}{1999}).

\bibitem[{\citenamefont{Reif}(1965)}]{enskog}
\bibinfo{author}{\bibfnamefont{F.}~\bibnamefont{Reif}},
  \emph{\bibinfo{title}{Fundamentals of statistical and thermal physics}},
  McGraw-Hill series in fundamentals of physics
  (\bibinfo{publisher}{McGraw-Hill}, \bibinfo{year}{1965}).

\bibitem[{\citenamefont{Roberts et~al.}(1998)\citenamefont{Roberts, Claussen,
  Burke, Greene, Cornell, and Wieman}}]{jacobscatter}
\bibinfo{author}{\bibfnamefont{J.~L.} \bibnamefont{Roberts}},
  \bibinfo{author}{\bibfnamefont{N.~R.} \bibnamefont{Claussen}},
  \bibinfo{author}{\bibfnamefont{J.~P.} \bibnamefont{Burke}},
  \bibinfo{author}{\bibfnamefont{C.~H.} \bibnamefont{Greene}},
  \bibinfo{author}{\bibfnamefont{E.~A.} \bibnamefont{Cornell}},
  \bibnamefont{and} \bibinfo{author}{\bibfnamefont{C.~E.}
  \bibnamefont{Wieman}}, \bibinfo{journal}{Phys. Rev. Lett.}
  \textbf{\bibinfo{volume}{81}}, \bibinfo{pages}{5109} (\bibinfo{year}{1998}).

\bibitem[{\citenamefont{Luiten et~al.}(1996)\citenamefont{Luiten, Reynolds, and
  Walraven}}]{walraven}
\bibinfo{author}{\bibfnamefont{O.~J.} \bibnamefont{Luiten}},
  \bibinfo{author}{\bibfnamefont{M.~W.} \bibnamefont{Reynolds}},
  \bibnamefont{and} \bibinfo{author}{\bibfnamefont{J.~T.~M.}
  \bibnamefont{Walraven}}, \bibinfo{journal}{Phys. Rev. A}
  \textbf{\bibinfo{volume}{53}}, \bibinfo{pages}{381} (\bibinfo{year}{1996}).

\bibitem[{\citenamefont{Hung et~al.}(2008)\citenamefont{Hung, Zhang, Gemelke,
  and Chin}}]{chin2D}
\bibinfo{author}{\bibfnamefont{C.-L.} \bibnamefont{Hung}},
  \bibinfo{author}{\bibfnamefont{X.}~\bibnamefont{Zhang}},
  \bibinfo{author}{\bibfnamefont{N.}~\bibnamefont{Gemelke}}, \bibnamefont{and}
  \bibinfo{author}{\bibfnamefont{C.}~\bibnamefont{Chin}},
  \bibinfo{journal}{Physical Review A} \textbf{\bibinfo{volume}{78}},
  \bibinfo{pages}{011604} (\bibinfo{year}{2008}).

\bibitem[{\citenamefont{Li et~al.}(2008)\citenamefont{Li, Alyabyshev, and
  Krems}}]{li2008}
\bibinfo{author}{\bibfnamefont{Z.}~\bibnamefont{Li}},
  \bibinfo{author}{\bibfnamefont{S.~V.} \bibnamefont{Alyabyshev}},
  \bibnamefont{and} \bibinfo{author}{\bibfnamefont{R.~V.} \bibnamefont{Krems}},
  \bibinfo{journal}{Phys. Rev. Lett.} \textbf{\bibinfo{volume}{100}},
  \bibinfo{pages}{073202} (\bibinfo{year}{2008}).

\bibitem[{\citenamefont{Kanjilal and Blume}(2004)}]{pwaveH}
\bibinfo{author}{\bibfnamefont{K.}~\bibnamefont{Kanjilal}} \bibnamefont{and}
  \bibinfo{author}{\bibfnamefont{D.}~\bibnamefont{Blume}},
  \bibinfo{journal}{Phys. Rev. A} \textbf{\bibinfo{volume}{70}},
  \bibinfo{pages}{042709} (\bibinfo{year}{2004}).

\bibitem[{\citenamefont{Swallows et~al.}(2011)\citenamefont{Swallows, Bishof,
  Lin, Blatt, Martin, Rey, and Ye}}]{anascience2011}
\bibinfo{author}{\bibfnamefont{M.~D.} \bibnamefont{Swallows}},
  \bibinfo{author}{\bibfnamefont{M.}~\bibnamefont{Bishof}},
  \bibinfo{author}{\bibfnamefont{Y.}~\bibnamefont{Lin}},
  \bibinfo{author}{\bibfnamefont{S.}~\bibnamefont{Blatt}},
  \bibinfo{author}{\bibfnamefont{M.~J.} \bibnamefont{Martin}},
  \bibinfo{author}{\bibfnamefont{A.~M.} \bibnamefont{Rey}}, \bibnamefont{and}
  \bibinfo{author}{\bibfnamefont{J.}~\bibnamefont{Ye}},
  \bibinfo{journal}{science} \textbf{\bibinfo{volume}{331}},
  \bibinfo{pages}{1043} (\bibinfo{year}{2011}).

\bibitem[{\citenamefont{Martin et~al.}(2013)\citenamefont{Martin, Bishof,
  Swallows, Zhang, Benko, von Stecher, Gorshkov, Rey, and Ye}}]{sr2013}
\bibinfo{author}{\bibfnamefont{M.~J.} \bibnamefont{Martin}},
  \bibinfo{author}{\bibfnamefont{M.}~\bibnamefont{Bishof}},
  \bibinfo{author}{\bibfnamefont{M.~D.} \bibnamefont{Swallows}},
  \bibinfo{author}{\bibfnamefont{X.}~\bibnamefont{Zhang}},
  \bibinfo{author}{\bibfnamefont{C.}~\bibnamefont{Benko}},
  \bibinfo{author}{\bibfnamefont{J.}~\bibnamefont{von Stecher}},
  \bibinfo{author}{\bibfnamefont{A.~V.} \bibnamefont{Gorshkov}},
  \bibinfo{author}{\bibfnamefont{A.~M.} \bibnamefont{Rey}}, \bibnamefont{and}
  \bibinfo{author}{\bibfnamefont{J.}~\bibnamefont{Ye}},
  \bibinfo{journal}{Science} \textbf{\bibinfo{volume}{341}},
  \bibinfo{pages}{632} (\bibinfo{year}{2013}).

\bibitem[{\citenamefont{Soltan-Panahi et~al.}(2011)\citenamefont{Soltan-Panahi,
  L{\"u}hmann, Struck, Windpassinger, and Sengstock}}]{soltan2011}
\bibinfo{author}{\bibfnamefont{P.}~\bibnamefont{Soltan-Panahi}},
  \bibinfo{author}{\bibfnamefont{D.-S.} \bibnamefont{L{\"u}hmann}},
  \bibinfo{author}{\bibfnamefont{J.}~\bibnamefont{Struck}},
  \bibinfo{author}{\bibfnamefont{P.}~\bibnamefont{Windpassinger}},
  \bibnamefont{and}
  \bibinfo{author}{\bibfnamefont{K.}~\bibnamefont{Sengstock}},
  \bibinfo{journal}{Nature Physics} \textbf{\bibinfo{volume}{8}},
  \bibinfo{pages}{71} (\bibinfo{year}{2011}).

\bibitem[{\citenamefont{Mark et~al.}(2011)\citenamefont{Mark, Haller, Lauber,
  Danzl, Daley, and N\"agerl}}]{mark2011}
\bibinfo{author}{\bibfnamefont{M.~J.} \bibnamefont{Mark}},
  \bibinfo{author}{\bibfnamefont{E.}~\bibnamefont{Haller}},
  \bibinfo{author}{\bibfnamefont{K.}~\bibnamefont{Lauber}},
  \bibinfo{author}{\bibfnamefont{J.~G.} \bibnamefont{Danzl}},
  \bibinfo{author}{\bibfnamefont{A.~J.} \bibnamefont{Daley}}, \bibnamefont{and}
  \bibinfo{author}{\bibfnamefont{H.-C.} \bibnamefont{N\"agerl}},
  \bibinfo{journal}{Phys. Rev. Lett.} \textbf{\bibinfo{volume}{107}},
  \bibinfo{pages}{175301} (\bibinfo{year}{2011}).

\bibitem[{\citenamefont{Will et~al.}(2010)\citenamefont{Will, Best, Schneider,
  Hackerm{\"u}ller, L{\"u}hmann, and Bloch}}]{will2010}
\bibinfo{author}{\bibfnamefont{S.}~\bibnamefont{Will}},
  \bibinfo{author}{\bibfnamefont{T.}~\bibnamefont{Best}},
  \bibinfo{author}{\bibfnamefont{U.}~\bibnamefont{Schneider}},
  \bibinfo{author}{\bibfnamefont{L.}~\bibnamefont{Hackerm{\"u}ller}},
  \bibinfo{author}{\bibfnamefont{D.-S.} \bibnamefont{L{\"u}hmann}},
  \bibnamefont{and} \bibinfo{author}{\bibfnamefont{I.}~\bibnamefont{Bloch}},
  \bibinfo{journal}{Nature} \textbf{\bibinfo{volume}{465}},
  \bibinfo{pages}{197} (\bibinfo{year}{2010}).

\bibitem[{\citenamefont{Lapidus}(1982)}]{lapidus1982}
\bibinfo{author}{\bibfnamefont{I.~R.} \bibnamefont{Lapidus}},
  \bibinfo{journal}{American Journal of Physics} \textbf{\bibinfo{volume}{50}},
  \bibinfo{pages}{45} (\bibinfo{year}{1982}).

\bibitem[{\citenamefont{Wu and Foot}(1996)}]{huang1996}
\bibinfo{author}{\bibfnamefont{H.}~\bibnamefont{Wu}} \bibnamefont{and}
  \bibinfo{author}{\bibfnamefont{C.~J.} \bibnamefont{Foot}},
  \bibinfo{journal}{Journal of Physics B: Atomic, Molecular and Optical
  Physics} \textbf{\bibinfo{volume}{29}}, \bibinfo{pages}{L321}
  (\bibinfo{year}{1996}).

\bibitem[{\citenamefont{Wu et~al.}(1997)\citenamefont{Wu, Arimondo, and
  Foot}}]{huang1997}
\bibinfo{author}{\bibfnamefont{H.}~\bibnamefont{Wu}},
  \bibinfo{author}{\bibfnamefont{E.}~\bibnamefont{Arimondo}}, \bibnamefont{and}
  \bibinfo{author}{\bibfnamefont{C.~J.} \bibnamefont{Foot}},
  \bibinfo{journal}{Phys. Rev. A} \textbf{\bibinfo{volume}{56}},
  \bibinfo{pages}{560} (\bibinfo{year}{1997}).

\bibitem[{\citenamefont{Holland et~al.}(2000)\citenamefont{Holland, DeMarco,
  and Jin}}]{tbfermi}
\bibinfo{author}{\bibfnamefont{M.~J.} \bibnamefont{Holland}},
  \bibinfo{author}{\bibfnamefont{B.}~\bibnamefont{DeMarco}}, \bibnamefont{and}
  \bibinfo{author}{\bibfnamefont{D.~S.} \bibnamefont{Jin}},
  \bibinfo{journal}{Phys. Rev. A} \textbf{\bibinfo{volume}{61}},
  \bibinfo{pages}{053610} (\bibinfo{year}{2000}).

\end{thebibliography}
\end{document}